\documentclass[11pt,a4paper]{article}

\usepackage{amsthm}
\usepackage{amsmath}
\usepackage{natbib}
\usepackage[colorlinks,citecolor=blue,urlcolor=blue,filecolor=blue,backref=page]{hyperref}
\usepackage{graphicx}
\usepackage{enumitem}
\usepackage{verbatim}
\usepackage{graphics}
\usepackage{color}
\usepackage{caption}
\usepackage{comment}
\usepackage{amssymb}
\usepackage{subcaption}
\usepackage{tikz}
\usepackage{algorithm}
\usepackage{multirow}
\usepackage{mathtools}
\usepackage[a4paper, lmargin=2cm, rmargin=2cm, bottom=3cm, top=2cm]{geometry}

\title{A Bayesian approach for determining player abilities in football}
\author{Gavin A.~Whitaker$^{1,2}$\thanks{email: \texttt{gavin.whitaker@ucl.ac.uk}} \and\ Ricardo Silva$^{1,3}$ \and\ Daniel Edwards$^2$ \and\ Ioannis Kosmidis$^{3,4}$}
\date{\small $^{1}$Department of Statistical Science, University College London, London, WC1E 6BT \\
$^{2}$Stratagem Technologies, 19 Eastbourne Terrace, London, W2 6LG \\
$^{3}$The Alan Turing Institute, 96 Euston Road, London, NW1 2DB\\
$^{4}$Department of Statistics, University of Warwick, Coventry, CV4 7AL}

\newcommand{\indep}{\overset{indep}{\sim }}
\newcommand{\expect}{\textrm{E}}
\newcommand{\elbo}{\textrm{ELBO}}

\begin{document}

\maketitle
\begin{abstract}
We consider the task of determining a football player's 
ability for a given event type, for example, scoring a goal. We propose 
an interpretable Bayesian model which is fit using 
variational inference methods. We implement a Poisson model to capture  
occurrences of event types, from which we infer player abilities. Our 
approach also allows the visualisation of differences between players, 
for a specific ability, through the marginal posterior variational 
densities. We then use these inferred player abilities to extend the 
Bayesian hierarchical model of \cite{baio_2010} which captures a 
team's scoring rate (the rate at which they score goals). We apply the 
resulting scheme to the English Premier League, capturing player 
abilities over the 2013/2014 season, before using output from the  
hierarchical model to predict whether over or under 2.5 goals will be 
scored in a given game in the 2014/2015 season. This validates 
our model as a way of providing insights into team formation and the 
individual success of sports teams.
\end{abstract}


\section{Introduction} \label{intro}

Our goal is to infer the ability of those players 
who play in the English Premier League. The Premier League is an 
annual football league established in 1992 and is the most 
watched football league in the world \citep{yueh_2014, curley_2016}. 
It is made up of 20 teams, who over the course of a season play every 
other team twice (both home and away), giving a total of 380 games 
each year. It is the top division of English football, and every year 
the bottom 3 teams are relegated to be replaced by 3 teams from the 
next division down (the Championship). In recent times the Premier 
League has also become known as the richest league in the world 
\citep{deloitte_2016} through both foreign investment and a 
lucrative deal for television rights \citep{cave_2016, rumsby_2016, bbc_2015}. 
Whilst there is growing financial competition from China, the Premier 
league arguably still attracts some of the best players in the world. 
Staying in the Premier league (by avoiding relegation) is worth a 
large amount of money, therefore teams are looking for any advantage 
when accessing a player's ability to ensure they sign the best players.
With large sums of money spent to buy/transfer these 
players it is natural to ask, ``How good are they at a specific skill, 
for example, passing a ball, scoring a goal or making a tackle?'' 
Here we present a method to access such ability, whilst quantifying 
the uncertainty about it for any given player.

The statistical modelling of sports has become a topic of increasing 
interest in recent times, as more data is collected through automation, 
coupled with a heightened interest in the outcome of these 
sports manifested by events such as the continuous rise of online betting. Football is 
providing an area of rich research, with the ability to capture the 
goals scored in a match being of particular interest. \cite{reep_1971} 
used a negative binomial distribution to model the aggregate goal 
counts, before \cite{maher_1982} used independent Poisson distributions 
to capture the goals scored by competing teams on a game by game basis. 
\cite{dixon_1997} also used the Poisson distribution to model scores, 
however they departed from the assumption of independence; the model 
is extended in \cite{dixon_1998}. The model of \cite{dixon_1997} is 
also built upon in \cite{karlis_2000, karlis_2003} who inflate the 
probability of a draw. \cite{baio_2010} consider this model in the 
Bayesian paradigm, implementing a Bayesian hierarchical model for 
goals scored by each team in a match. A Weibull count model for scores 
is explored by \cite{boshnakov_2017}. \cite{groll_2018} use a bivariate 
Poisson model for goals scored by national teams competing in the 2016 
UEFA European football championship. Other works to investigate the 
modelling of football scores include \cite{lee_1997}, 
\cite{joseph_2006} and \cite{karlis_2009}.

A player performance rating system (the EA Sports Player Performance 
Index) was developed by \cite{mchale_2012}. The rating system was  
developed in conjunction with the English Premier League, the English 
Football League, Football DataCo and the Press Association, and aims 
to represent a player's worth in a single number. There is some debate 
within the football community on the weightings derived in the paper, 
and as \cite{mchale_2012} point out, the players who play for the best 
teams lead the index. There is also some questions raised as to whether 
reducing the rating to a single number (whilst easy to understand) 
masks a player's ability in a certain skill, whether good or bad. 
Finally, as mentioned by the authors, the rating system does not 
handle those players who sustain injuries (and therefore have little 
playing time) well. 
TrueSkill\texttrademark \citep{herbrich_2007} has also been proposed 
as a method to rate (and rank) a player's ability at a skill; initially 
this was the ability to play Halo 2 online, a video game in the Microsoft Xbox network. 
However, one shortcoming  of TrueSkill is that it cannot distinguish between two players who 
always play on the same team. Whilst this is not a problem for online 
gaming where teams often change, or players play with different groups of 
friends, in football this causes a more substantial issue. Two football 
players regularly play on the same team, meaning they would be 
indistinguishable in their ability. If TrueSkill were used for the applications in  
this paper, this problem would be observed frequently for the dataset 
considered. A second version, TrueSkill2, is presented 
by \cite{minka_2018} which incorporates additional statistics 
to address this issue. In football it is crucial to be able to distinguish 
between two given players, even if they always play on the same team. 
\cite{mchale_2014} attempt to identify the goal 
scoring ability of players. Spatial methods to capture a team's 
style and behaviour are explored in \cite{lucey_2013}, 
\cite{bialkowski_2014} and \cite{bojinov_2016}, with \cite{whitaker_2018} 
applying a spatial model to attacking events only. Here, our interest 
lies first in defining player ability whilst addressing some of the issues 
raised by \cite{mchale_2012}, before attempting to capture the 
goals scored in a game taking into account these abilities.
More specifically, we wish to calculate each player's ability at a skill, 
for example, scoring a goal, and hope to derive a meaningful ranking of players 
from these perceived abilities. From this ranking we can answer the 
hotly debated questions of football, such as, ``Who is the 
best goalscorer?'' Such questions are often asked in sports, 
and frequently the answers are given from little more than subjective 
views. We believe the methods presented in this paper offer a more 
evidence based approach. We model the goals scored in a game 
(and ultimately make predictions) incorporating these abilities.
This is of interest not only to football enthusiasts, and stakeholders in 
football clubs, but also to all industries engaging in sports trading and sports advertising.

To fit our proposed model (and infer player abilities) we appeal to variational inference (VI) 
methods, an alternative to Markov chain Monte Carlo (MCMC) 
sampling which can be advantageous to use when datasets are large 
and/or models have high complexity. Popularised in the machine learning 
literature \citep{jordan_1999, wainwright_2008}, 
VI transforms the problem of approximate posterior inference into an 
optimisation problem, making it easier to scale to large data than MCMC. 
Some application areas and indicative references where VI has been used include 
sports \citep{kitani_2011, ruiz_2015, franks_2015}, 
computational biology \citep{carbonetto_2012, raj_2014}  
and computer vision \citep{blei_2006, sudderth_2009, du_2009}.  
For a discussion on VI techniques as a whole see \cite{blei_2017} and 
the references therein.

The remainder of the paper is organised as follows. The data is 
presented in Section~\ref{data}. In Section~\ref{bayes} we outline our 
model to define player abilities before discussing a variational 
inference approach; we finish the section by offering our extension 
to the Bayesian hierarchical model of~\cite{baio_2010}. Applications 
are considered in Section~\ref{app} and a discussion is provided in 
Section~\ref{disc}.

\section{The data} \label{data}

The data available to us is a collection of touch-by-touch data, which 
records every touch in a given game, noting the time, team, player, 
type of event and outcome. 
The data covers the 2013/2014 and 2014/2015 
English Premier League seasons and consists of roughly 
1.2 million events in total, which equates to approximately 1600 for 
each game. There are 39 event types in the dataset which we list 
in Table~\ref{eventtype}. The nature of most of these event types is 
self-explanatory, that is, Goal indicates that a player scored a goal 
at that event time. Throughout this paper we will mainly concern 
ourselves with event types which are self-evident, and will define the 
more subtle event types when needed. 

\begin{table} 
\caption{Event types contained within the data.} \label{eventtype}
\centering
\begin{tabular}{llll}
\hline
Stop & Control & Disruption & Miscellanea \\
\hline
 Card & Aerial & BlockedPass & CornerAwarded\\
   End & BallRecovery & Challenge & CrossNotClaimed\\
     FormationChange & BallTouch & Claim & KeeperSweeper\\
       FormationSet & ChanceMissed & Clearance & ShieldBallOpp\\
         OffsideGiven & Dispossessed & Interception & \\
           PenaltyFaced & Error & KeeperPickup & \\
             Start & Foul & OffsideProvoked & \\
               SubstitutionOff & Goal & Punch & \\
                 SubstitutionOn & GoodSkill & Save & \\
                    & MissedShots & Smother & \\
                      & OffsidePass & Tackle & \\
                        & Pass &  &  \\
                          & SavedShot & &  \\
                            & ShotOnPost &  & \\
                               & TakeOn &  & \\
\hline
\end{tabular}
\end{table} 

We can split the event types into 4 categories.  
\begin{enumerate}
\item \textbf{Stop:} An event corresponding to a stoppage in play such 
as a substitution or offside decision.
\item \textbf{Control:} An event where a team is perceived to be in 
control of the ball, these are mainly seen as attacking events.
\item \textbf{Disruption:} An event where a team is perceived to be 
disrupting the current play within a game, these can generally be seen 
as defensive events.
\item \textbf{Miscellanea:} These events could be classified in any of the 
other three categories. 
\end{enumerate}
In this paper we are interested in the events which correspond to a 
player during active game-play, hence we remove Stop events from the 
data when considering player abilities. We note however, that we use 
SubstitutionOn and SubstitutionOff when calculating a player's playing 
time in a game, although we do not model abilities for these events. 
We focus on those event types categorised as either 
Control or Disruption, that is, when a team is attempting to score a
goal and when a team is attempting to stop the opposition from scoring 
a goal respectively. 

\begin{figure}
      \centering
      \includegraphics[scale=0.8]{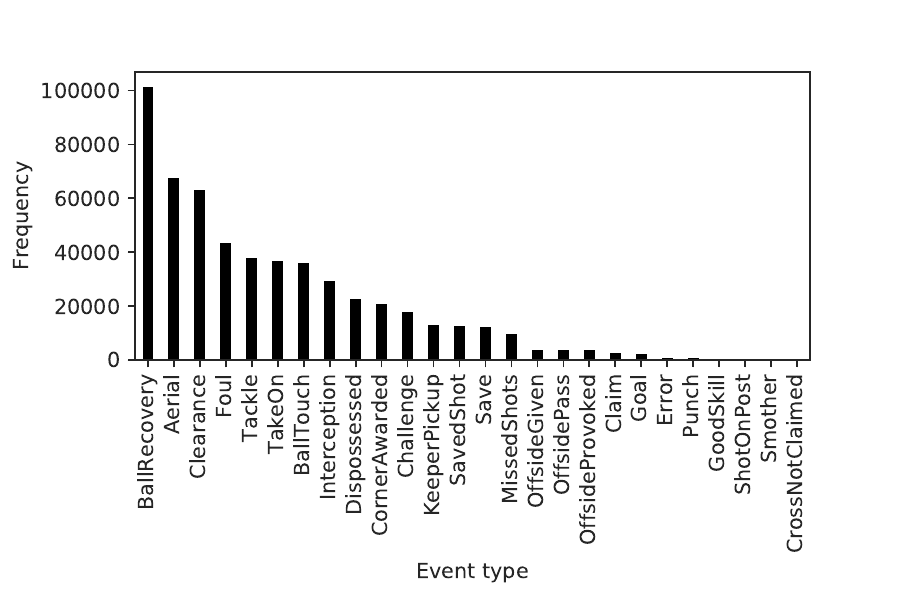}
      \caption{Frequency of each event type observed in the 2013/2014 
      and 2014/2015 English Premier League seasons. 
      The event type Pass is removed for clarity as it occurs with a 
      ratio of approximately 10:1 over BallRecovery.} \label{typefig}
\end{figure}

It should be noted that OffsideGiven is the inverse of OffsideProvoked
and as such we remove one of these events from the data. Henceforth, it 
is assumed that the event type OffsideGiven is removed from the 
data, rewarding the defensive side for provoking an offside through 
OffsideProvoked. The frequency of each event type (after removing Pass) 
during the 2013/2014 and 2014/2015 English Premier League seasons 
is shown in Figure~\ref{typefig}. Pass dominates the data over all other 
event types recorded, with a ratio of approximately 10:1 to BallRecovery, 
and hence is removed for clarity. This is not surprising given the make 
up of a football match (where teams mainly pass the ball).

In determining a player's ability for a given event type we make the 
assumption that the more times a player is involved the better they 
are at that event type. For example, a player who makes more passes 
than another player is assumed to be the better passer. On this basis, 
we can transform the data 
to represent the number of each event type each player is involved in at 
a game by game level. This count data is illustrated in 
Table~\ref{tab-count}. It is to this data which the methods of 
Section~\ref{bayes} will be applied.

\begin{table} 
\caption{A section of the count data.} \label{tab-count}
\centering
\begin{tabular}{ccc|cccc}
\hline
 & & & \multicolumn{4}{c}{Count for each event type} \\
game id &	player id & team id &	Goal & Pass &	Tackle &	$\cdots$\\
\hline
1483412  &	17 &	663 & 0 & 97 & 3 &    \\
1483412 &	3817 &	663 & 0 & 37 & 3 &   	\\
1483412 &	4574 &	663 & 0 & 73 & 3 &   	\\
\vdots & \vdots &\vdots &\vdots &\vdots &\vdots & \\[1.1ex]
1483831 &	10136 &	676 & 1 & 36 & 4 &    \\
1483831 &	12267 &	676 & 0 & 45 & 0 &   	\\
1483831 &	12378 &	676 & 0 & 52 & 2 &   	\\
\vdots & \vdots &\vdots &\vdots &\vdots &\vdots & \\
\hline
\end{tabular}
\end{table}

\section{Model} \label{bayes}

Consider the case where we have $K$ matches, numbered $k=1,\ldots,K$. 
We denote the set of teams in game $k$ as $T_k$, with $T_k^H$ and 
$T_k^A$ representing the home and away teams respectively. Hence, 
$T_k = \{T_k^H, T_k^A\}$. We take $P$ to be the set of all players who 
feature in the dataset, and $P_k^j\subset P$ to be the subset of players 
who play for team~$j$ in game $k$. We may want to consider how 
players' abilities over different event types interact. For this, we 
group event types to create meaningful interactions. In grouping events 
there is an underlying assumption about the independence between the 
events that make up a grouped event. To the authors knowledge there is 
no evidence in the existing literature of interactions which are clearly 
detectable from data (a question we believe would make a very interesting 
area of future research). For simplicity, 
we describe the model for a single pair of event types which are deemed 
to interact, for example, Pass and Interception, and we denote these event 
types $e_1$ and $e_2$, with $E=\{e_1, e_2\}$.

Taking $X_{i,k}^{e}$ as the number of occurrences (counts) of event 
type $e$, by player $i$ (who plays for team~$j$), in match $k$, we assume that 
\begin{equation} \label{posmod}
X_{i,k}^{e}\sim Poisson\left(\eta_{i,k}^{e}\tau_{i,k} \right), 
\end{equation}
where
\begin{equation} \label{eta}
\eta_{i,k}^{e} = \exp\left\{\Delta_i^{e} 
                          + \tau_{i,k}\left(\lambda_1^e\sum_{i'\in P_k^j}\Delta_{i'}^{e}  
                          - \lambda_2^e\sum_{i'\in P_k^{T_k\setminus j}}\Delta_{i'}^{E\setminus e}\right)  
                          + \left(\delta_{T_k^H,j}\right)\gamma^{\,e} \right\},
\end{equation}
$\delta_{r,s}$ is the Kronecker delta and $\tau_{i,k}$ is the 
fraction of time player $i$ (playing for team $j$) spent on the pitch 
in match $k$, with $\tau_{i,k}\in[0, 1]$. Specifically, if a player plays 
60 minutes of a 90 minute match then $\tau_{i,k}=2/3$. The home effect 
is represented by $\gamma^{\,e}$. The home effect reflects the (supposed) 
advantage the home team has over the away team in event type $e$. The 
$\Delta_i^{e}$ represent the (latent) ability of each player for a 
specific event type, where we let $\Delta$ be the vector of all 
players abilities. The impact of a player's own team on the number of 
occurrences is captured through $\lambda_1^e$, with $\lambda_2^e$ 
describing the opposition's ability to stop the player in that event 
type. Whilst the inclusion 
of $\tau_{i,k}$ in \eqref{posmod} is obvious, its inclusion in \eqref{eta} 
is more nuanced. Here, across different games with the same player, we aim 
to capture variability in rate $\eta_{i,k}^{e}$ as team strategy varies 
from game to game: a change of time played by player $i$ explains other 
unobserved features of the game meant to affect rate $\eta_{i,k}^e$ 
directly. A version of~\eqref{eta} was initially considered where a 
player would have their own individual parameter governing their ability to 
interact with their team in an event type and how put off they would 
be by the opposition. We found that this led to a model with too many 
parameters for succinct consideration and fitting. Expert football 
analysts suggested that a time a player played in a given game would act as a good 
proxy in the absence of a specific parameter. The inclusion of $\tau_{i,k}$ in 
\eqref{eta} is discussed further in Section~\ref{ability}. As $\tau_{i,k}\in[0, 1]$ 
a player therefore receives a proportion of their own teams ability 
(and opposition's ability to stop) determined by the time they play in 
a given game. For identifiability purposes, we impose the constraint that the 
$\lambda$s must be positive. Figure~\ref{pic-model} illustrates 
the model for one game, allowing for some abuse in notation, 
and where we assume each team consists of 11 players only (that is, 
we ignore substitutions) and suppress the time dependence ($\tau$). 
From \eqref{posmod} and \eqref{eta}, the log-likelihood is given by
\begin{equation}
\ell = \sum_{e\in E}\sum_{k=1}^K\sum_{j\in T_k}\sum_{i\in P_k^j} X_{i,k}^{e} \log{\left(\eta_{i,k}^{e}\tau_{i,k} \right)} - \eta_{i,k}^{e}\tau_{i,k} - \log{\left(X_{i,k}^{e}\,! \right)}. \label{llike}
\end{equation}

\begin{figure}
\centering
\begin{tikzpicture}[scale=5,>=latex]
     \node[draw,circle,fill=black,inner sep=0.5mm, label=above:{\large $\Delta_1^{e_1}$}]
         (xo0h) at (0,1) {};
     \node[draw,circle,fill=black,inner sep=0.5mm, label=above:{\large $\Delta_2^{e_1}$}]
         (xo1h) at (0.3,1) {};
     \node at (0.5, 1) {{$\ldots$}};   
     \node[draw,circle,fill=black,inner sep=0.5mm, label=above:{\large $\Delta_{11}^{e_1}$}]
         (xo2h) at (0.7,1) {};
         
     \node[draw,circle,fill=black,inner sep=0.5mm, label=above:{\large $\Delta_{12}^{e_2}$}]
         (xo0a) at (1.4,1) {};
     \node[draw,circle,fill=black,inner sep=0.5mm, label=above:{\large $\Delta_{13}^{e_2}$}]
         (xo1a) at (1.7,1) {};
     \node at (1.9, 1) {{$\ldots$}};   
     \node[draw,circle,fill=black,inner sep=0.5mm, label=above:{\large $\Delta_{22}^{e_2}$}]
         (xo2a) at (2.1,1) {};
         
     \node[draw,circle,fill=black,inner sep=0.5mm, label={[yshift=-0.9cm]\large $\log\left(\eta_1^{e_1}\right)$}]
         (xu0) at (0,0) {};

    \node[draw,circle]
         (hsum) at (0.3,0.5) {\large $\displaystyle\lambda_1\sum_{i=1}^{11}\Delta_i^{e_1}$};

    \node[draw,circle]
         (asum) at (1.7,0.5) {\large $\displaystyle-\lambda_2\sum_{i=12}^{22}\Delta_i^{e_2}$};
         
     \node[draw,circle]
         (gam) at (0.7,0.3) {\large $\gamma^{\;e_1}$};

      \draw (xo0h) edge[out=240,in=120,->]  (xu0) ;
      
      \draw (xo0h) edge[->]  (hsum) ;  
      \draw (xo1h) edge[->]  (hsum) ;
      \draw (xo2h) edge[->]  (hsum) ;
      
      \draw (xo0a) edge[->]  (asum) ;  
      \draw (xo1a) edge[->]  (asum) ;
      \draw (xo2a) edge[->]  (asum) ;
      
      \draw (hsum) edge[->]  (xu0) ;  
      \draw (gam) edge[->]  (xu0) ;
      \draw (asum) edge[out=220,in=-20,->]  (xu0) ;

\end{tikzpicture}
\caption{Pictorial representation of the model for one game. For ease we assume that only 11 players play for each team (that is, we ignore substitutions) and suppress the time dependence ($\tau$).} \label{pic-model}
\end{figure}

Interest lies in estimating this model using a Bayesian approach. We 
put independent Gaussian priors over all abilities, whilst treating 
the remaining unknown parameters as hyperparameters to be fitted 
by the marginal likelihood function. Given the size of the data and 
the number of parameters needing to be estimated to fit 
equation~\eqref{llike} we appeal to variational inference techniques, 
which are the subject of the next section.

    \subsection{Variational inference} \label{var-inf}

In contrast to some other techniques for Bayesian inference, such as 
MCMC, in variational inference~(VI) we specify a variational family of densities over 
latent variables. We introduce the basic idea with a general notation, using $x$ to denote observed data and $\nu$ the set of latent variables, here assumed to be continuous. In VI we aim to find the best candidate 
approximation, $q^*(\nu)$, to minimise the Kullback-Leibler~(KL) 
divergence to the posterior $\pi(\nu\vert x)$ within a suitable space $\mathcal Q$ of density functions
\[
q^*(\nu) = \underset{q\left(\nu\right) \in \mathcal Q}{\textrm{argmin KL}}\left\{q(\nu)\vert\vert\pi(\nu\vert x) \right\}.
\]
Unfortunately, due to the analytic intractability of the posterior 
distribution the KL divergence is not available in closed (analytic) 
form. However, it is possible to maximise the evidence lower bound~(ELBO). 
The ELBO is the expectation of the joint density under the 
approximation plus the entropy of the variational density and is 
given by
\begin{equation}\label{elbo}
\elbo\left\{q\left(\nu\right)\right\} = \expect_{q\left(\nu\right)}\left[\log\left\{\pi\left(\nu,x\right)\right\}\right] - \expect_{q\left(\nu\right)}\left[\log\left\{q\left(\nu\right)\right\}\right].
\end{equation}
The ELBO is the equivalent of the negative KL divergence up to the 
constant $\log\{\pi(x)\}$, and from \cite{jordan_1999} and 
\cite{bishop_2006} we know that by maximising the ELBO we minimise 
the KL divergence. 

In performing VI, assumptions must be made about the variational 
family $\mathcal Q$ to which $q(\nu)$ belongs. Here we consider the \emph{mean-field 
variational family}, in which the latent variables are assumed to be 
mutually independent. Moreover, each latent variable $\nu_r$ is 
governed by its own variational parameters~$\phi_r$. They determine 
$\nu_r$'s variational factor, the density $q(\nu_r\vert \phi_r)$. 
Specifically, for $R$ latent variables, we have that
\begin{equation}\label{meanfield}
q\left(\nu \right) = \prod_{r=1}^Rq\left(\nu_r\vert\phi_r\right).
\end{equation} 
We note that the complexity of the variational family determines the 
complexity of the optimisation, and hence impacts the computational 
cost of any VI approach. In general, it is possible to impose any 
graphical structure on $q(\nu)$; a fully general 
graphical approach leads to structured variational inference, see 
\cite{saul_1996}. Furthermore, the data ($x$) does not feature in 
equation~\eqref{meanfield}, meaning the variational family is not a 
model of the observed data. It is in fact the ELBO which connects the 
variational density, $q(\nu)$, to the data and the model.

For the model outlined at the beginning of this section, let 
$\nu=\Delta$, and set
\begin{equation}\label{qDelta_single}
q\left(\Delta_i^{e}\vert\phi_i^{e} \right)\sim N\left(\mu_{\Delta_i^{e}}, \sigma_{\Delta_i^{e}}^2 \right). 
\end{equation}
Our aim is to find suitable candidate values for the variational 
parameters
\[
\phi_i^{e}=\left(\mu_{\Delta_i^{e}},\sigma_{\Delta_i^{e}}\right)',\qquad\forall i ,\forall e,
\]
where $'$ denotes the transpose. We take
\begin{equation}\label{qDelta}
q\left(\Delta \right) = \prod_{e\in E}\prod_{j\in T}\prod_{i\in P^j} q\left(\Delta_i^{e}\vert\phi_i^{e} \right), 
\end{equation}
where $T$ is the set of all teams and $P^j$ are the players who play 
for team $j$. Finally we take $\psi=~(\lambda_1^{e_1}, \lambda_2^{e_1}, \gamma^{\,e_1},\lambda_1^{e_2}, \lambda_2^{e_2}, \gamma^{\,e_2})'$ 
to be fixed parameters, and assume each $\Delta_i^{e}$ follows a 
$N(m,s^2)$ prior distribution, fully specifying the model given by 
equations~\eqref{posmod}--\eqref{llike}. Thus, the ELBO \eqref{elbo} is 
given by 
\begin{align}
\elbo\left\{q\left(\Delta\right)\right\} &= \sum_{e\in E}\sum_{k=1}^K\sum_{j\in T_k}\sum_{i\in P_k^j} \expect_{q\left(\Delta_i^{e}\right)}\left[\log\left\{\pi\left(\Delta_i^{e},\phi_i^{e},\psi,x\right)\right\}\right] - \expect_{q\left(\Delta_i^{e}\right)}\left[\log\left\{q\left(\Delta_i^{e}\vert\phi_i^{e}\right)\right\}\right] \nonumber \\
& = \sum_{e\in E}\sum_{k=1}^K\sum_{j\in T_k}\sum_{i\in P_k^j} \expect_{q\left(\Delta_i^{e}\right)}\left[\log\left\{\pi\left(\Delta_i^{e}\right)\right\}\right] + \expect_{q\left(\Delta_i^{e}\right)}\left[\log\left\{\pi\left(x\vert\Delta_i^{e},\phi_i^{e},\psi\right)\right\}\right] \nonumber\\
&\qquad\qquad\qquad\qquad\quad- \expect_{q\left(\Delta_i^{e}\right)}\left[\log\left\{q\left(\Delta_i^{e}\vert\phi_i^{e}\right)\right\}\right]. \label{model_elbo}
\end{align}
The above is available in closed-form (see Appendix~\ref{app_elbo} for the explicit expressions). Whilst in classical VI optimisation is done by coordinate ascent, where the optimal solution for each $\phi_i^e$ can be found in closed form by fixing all other free parameters, it is widely accepted today that generic gradient-based optimisers are better suited. In particular, with the popularisation of automatic differentiation tools such as the \verb+Python+ package \verb+autograd+ \citep{maclaurin_2015}, programming requirements are minimal: it suffices to implement only \eqref{model_elbo}, as given in Appendix~\ref{app_elbo}, and pass this function call to a gradient-based optimisation library that can internally make use of a tool such as \verb+autograd+. Whilst libraries such as the Automatic Differentiation Variational Inference (ADVI) engine by \cite{kucukelbir_2016} are available, they are particularly useful when the ELBO cannot be computed analytically. In our case, we found it easier to implement \eqref{model_elbo} in native \verb+Python+ along with a call to a standard optimiser (ADAM) coupled with \verb+autograd+.

    \subsection{Hierarchical model} \label{hier}

Building on the methods of Section~\ref{var-inf}, we wish to 
discover whether the inferred $\Delta$s have any impact on our 
ability to predict the goals scored in a football match. This provides an indirect
and objective validation that our estimated latent abilities are capturing information concerning a player's performance. As a 
baseline model we consider the work of \cite{baio_2010}, who present 
the model of \cite{karlis_2003} in a Bayesian framework. The model has 
close ties with the models in \cite{dixon_1997}, \cite{lee_1997} and \cite{karlis_2000}, which have 
all previously been used to predict football scores. We first briefly 
outline the model of \cite{baio_2010} before offering our extension 
to include the imputed $\Delta$s.

The model is a Poisson-log normal model; see for example 
\cite{aitchison_1989}, \cite{chib_2001} and \cite{tunaru_2002} (amongst 
others). For a particular game $k$ we let $y^k=~(y^k_h,y^k_a)'$ 
be the total number of goals scored, where $y^k_h$ is the number of 
goals scored by the home team, and $y^k_a$ the number by the away 
team. Inherently, we let $h$ denote the home team and $a$ the away 
team for the given game $k$. The goals of each team are modelled by 
independent Poisson distributions, such that 
\begin{equation}\label{hier_pois}
y_t^k\vert\theta_t\indep Poisson\left(\theta_t \right), \qquad t\in\{h,a\},
\end{equation} 
where 
\begin{align}
\log{\left(\theta_h\right)} &= \textrm{home} + \textrm{att}_h +\textrm{def}_a, \label{theta_h}\\ 
\log{\left(\theta_a\right)} &= \textrm{att}_a + \textrm{def}_h. \label{theta_a}
\end{align}
Each team has their own team-specific attack and defence ability, 
$\textrm{att}$ and $\textrm{def}$ respectively, which form the scoring 
intensities $(\theta_t)$ of the home and away teams. A home 
effect $(\textrm{home})$, which is assumed to be constant across all 
teams and across the time-span of the data is also included in the 
rate of the home team's goals. 

For identifiability we follow \cite{baio_2010} and \cite{karlis_2003}, 
and impose sum to zero constraints on the attack and defence 
parameters 
\[
\sum_{t\in T} \textrm{att}_t = 0 \qquad \textrm{and} \qquad\sum_{t\in T} \textrm{def}_t = 0,
\]
where $T$ is the set of all teams to feature in the dataset. 
Furthermore, the attack and defence parameters for each team are seen 
to be draws from a common distribution 
\[
\textrm{att}_t\sim N\left(\mu_{\textrm{att}}, \sigma_{\textrm{att}}^2\right) \qquad \textrm{and} \qquad \textrm{def}_t\sim N\left(\mu_{\textrm{def}}, \sigma_{\textrm{def}}^2\right).
\]
We follow the prior setup of \cite{baio_2010} and assume that 
$\textrm{home}$ follows a $N(0,100^2)$ distribution 
\emph{a priori}, with the hyper parameters having the priors 
\begin{alignat}{3}
&\mu_{\textrm{att}}\sim N\left(0,100^2\right), &&\qquad \mu_{\textrm{def}}\sim N\left(0,100^2\right), \nonumber\\
&\sigma_{\textrm{att}}\sim Inv\textrm{-}Gamma(0.1,0.1), &&\qquad \sigma_{\textrm{def}}\sim Inv\textrm{-}Gamma(0.1,0.1).\nonumber
\end{alignat}
A graphical representation of the model is 
given in Figure~\ref{pic-hierarchical}. 

\begin{figure}
\centering
\begin{tikzpicture}[scale=2,>=latex]

    \node[draw,circle]
         (mu_a) at (0,3) {\large $\mu_{\textrm{att}}$};
    \node[draw,circle]
         (sigma_a) at (1,3) {\large $\sigma_{\textrm{att}}$};
         
    \node[draw,circle]
         (mu_d) at (3,3) {\large $\mu_{\textrm{def}}$};
    \node[draw,circle]
         (sigma_d) at (4,3) {\large $\sigma_{\textrm{def}}$};
         
    \node[draw,circle]
         (home) at (-0.15,2) {\large $\textrm{home}$};         
    \node[draw,circle]
         (att_h) at (0.5,2) {\large $\textrm{att}_{h}$};
    \node[draw,circle]
         (def_a) at (1.1,2) {\large $\textrm{def}_{a}$};
         
    \node[draw,circle]
         (att_a) at (3.2,2) {\large $\textrm{att}_{a}$};
    \node[draw,circle]
         (def_h) at (3.8,2) {\large $\textrm{def}_{h}$};

    \node[draw,circle,dashed, black!70!]
         (delta_h) at (-0.95,1.6) {\large $f(q(\Delta))_h$};  
    \node[draw,circle,dashed, black!70!]
         (delta_a) at (4.53,1.6) {\large $f(q(\Delta))_a$}; 
                  
    \node[draw,circle]
         (theta_h) at (0.5,1.25) {\large $\theta_{h}$};
    \node[draw,circle]
         (theta_a) at (3.5,1.25) {\large $\theta_{a}$};         

    \node[draw,circle]
         (y_h) at (0.5,0.5) {\large $y_{h}$};
    \node[draw,circle]
         (y_a) at (3.5,0.5) {\large $y_{a}$}; 
      
      \draw (mu_a) edge[->]  (att_h) ; 
      \draw (mu_a) edge[->]  (att_a) ;  
      \draw (sigma_a) edge[->]  (att_h) ; 
      \draw (sigma_a) edge[->]  (att_a) ;
            
      \draw (mu_d) edge[->]  (def_h) ;
      \draw (mu_d) edge[->]  (def_a) ;
      \draw (sigma_d) edge[->]  (def_h) ; 
      \draw (sigma_d) edge[->]  (def_a) ;

      \draw (home) edge[->]  (theta_h) ;
      \draw (att_h) edge[->]  (theta_h) ;
      \draw (def_a) edge[->]  (theta_h) ;
      
      \draw (att_a) edge[->]  (theta_a) ;
      \draw (def_h) edge[->]  (theta_a) ; 
      
      \draw (delta_h) edge[dashed, ->, black!70!]  (theta_h) ;   
      \draw (delta_a) edge[dashed, ->, black!70!]  (theta_a) ;   
             
      \draw (theta_h) edge[->]  (y_h) ;
      \draw (theta_a) edge[->]  (y_a) ; 
         
\end{tikzpicture}
\caption{Pictorial representation of the Bayesian hierarchical model. Removing both $f(q(\Delta))_h$ and $f(q(\Delta))_a$ gives the baseline model of \cite{baio_2010}.}\label{pic-hierarchical}
\end{figure}

As an extension to the model of \cite{baio_2010} we propose to include information about the distribution of the latent $\Delta$s of Section~\ref{var-inf} in the scoring 
intensities of both the home and away teams. Explicitly \eqref{theta_h} 
and \eqref{theta_a} become
\begin{align}
\log{\left(\theta_h\right)} &= \textrm{home} + \textrm{att}_h +\textrm{def}_a + f\left(q(\Delta)\right)_h, \label{theta_h_ext}\\
\log{\left(\theta_a\right)} &= \textrm{att}_a + \textrm{def}_h + f\left(q(\Delta)\right)_a, \label{theta_a_ext}
\end{align}
where $f(q(\Delta))$ is to be determined. For a single pair of event 
types (as outlined at the start of this section), our choice 
for $f(q(\Delta))$ in this paper is
\begin{align}
f\left(q(\Delta)\right)_h &= \sum_{i\in I_k^{T_k^H}}\mu_{\Delta_i^{e}} - \sum_{i\in I_k^{T_k^A}}\mu_{\Delta_{i}^{E\setminus e}} \label{fdeltah}
\shortintertext{and}
f\left(q(\Delta)\right)_a &= \sum_{i\in I_k^{T_k^A}}\mu_{\Delta_{i}^{e}} - \sum_{i\in I_k^{T_k^H}}\mu_{\Delta_{i}^{E\setminus e}}, \label{fdeltaa}
\end{align}
with $I_k^j$ being the initial eleven players who start game $k$ for 
team $j$, and $\mu_{\Delta}$ being the mean of the marginal posterior 
variational densities. This extension is also illustrated in 
Figure~\ref{pic-hierarchical}. Through our empirical investigations we 
found little difference between trying different event types one by one 
or where $f(q(\Delta))$ is a summation over event types. Therefore, for 
ease, from this point forward we only consider the case in which 
$f(q(\Delta))$ is a summation over event types. The question around which 
specific event types should enter this summation is considered in 
Section~\ref{prediction}.

We fit both the baseline model of~\cite{baio_2010} and our extension 
using \verb+PyStan+~\citep{stan_2016}.

\section{Applications} \label{app}

Having outlined our approach to determine a player's ability in a 
given event type, and offered an extension to the model of 
\cite{baio_2010} to capture the goals scored in a specific game, we 
wish to test the proposed methods in real world scenarios. We 
therefore consider two applications. In the first we use data from the 
2013/2014 English Premier League to learn players abilities across the 
season as a whole for a number of event types, including the ability 
to score a goal. The second example concerns the number of goals 
observed in a given game, specifically we predict whether a 
certain number of goals will be scored (or not) in each game, 
offering validation against the betting market.

    \subsection{Determining a player's ability} \label{ability}

In this section we consider the touch-by-touch data described in 
Section~\ref{data} and consider data for the 2013/2014 English Premier 
League season only. We look to create an ordering of players 
abilities, from which we hope to extract meaning based on what we know 
of the season. We also have data on the amount of time each player 
spent on the pitch in each match and this information is factored in 
accordingly through $\tau_{i,k}$. The season consisted of 380 matches 
for the 20 team league, with 544 different players used during matches. 
The final league table is shown in Table~\ref{finaltable}. From this table,
we note that Manchester City and Liverpool were the teams who scored 
the most goals, with Chelsea conceding the least. These teams did well 
over the season and we expect players from these teams to have high 
abilities. The teams to do worst (and got relegated), were Norwich 
City, Fulham and Cardiff; we do not expect players from these teams to 
feature highly in any ordering created. A final note is that, in this 
season Manchester United underperformed (given past seasons) under 
new manager David Moyes. Whence, $k=1,\ldots,380$, $j\in T_k$ where 
$T_k$ consists of a subset of $\{1,\ldots,20\}$ and $i\in P_k^j$ where 
$P_k^j$ is a subset of $P=\{1,\ldots,544\}$.

\begin{table}[t]
\caption{Final league table for the 2013/2014 English Premier League. \emph{Pl} matches played, \emph{W}~matches won, \emph{D} matches drawn, \emph{L} matches lost, \emph{F} goals scored, \emph{A} goals conceded, {GD} goal difference ($\textrm{scored}-\textrm{conceded}$), \emph{Pts} final points total.} \label{finaltable}
\centering
\begin{tabular}{clcccccccc}
\hline
\multicolumn{10}{c}{2013/2014 English Premier League} \\
\hline 
Pos & Team & Pl &	W &	D &	L &	F &	A &	GD &	Pts \\
1 &	Manchester City & 	38 &	27 &	5 &	6 &	102 &	37 &	65 &	86 \\
2 &	Liverpool &	38 &	26 &	6 &	6 &	101 &	50 &	51 &	84 \\
3 &	Chelsea &	38 &	25 &	7 &	6 &	71 &	27 & 44 &	82 \\
4 &	Arsenal &	38 &	24 &	7 &	7 &	68 &	41 &	27 &	79 \\
\hline
5 &	Everton &	38 &	21 &	9 &	8 &	61 &	39 &	22 &	72\\
6 &	Tottenham Hotspur &	38 &	21 &	6 &	11 &	55 &	51 	& 4 &	69 \\
\hline
7 &	Manchester United &	38 &	19 &	7 &	12 &	64 &	43 &	21 &	64\\
8 &	Southampton &	38 &	15 &	11 &	12 &	54 &	46 &	8 &	56\\
9 &	Stoke City &	38 &	13 &	11 &	14 &	45 &	52 &	-7 &	50\\
10 &	Newcastle United &	38 &	15 &	4 &	19 &	43 &	59 &	-16 &	49\\
11 &	Crystal Palace &	38 	&13 &	6 &	19 &	33 &	48 &	-15 &	45\\
12 &	Swansea City &	38 &	11 &	9 &	18 &	54 &	54 &	0 &	42\\
13 &	West Ham United &	38 &	11 &	7 &	20 &	40 &	51 &	-11 &	40\\
14 &	Sunderland &	38 &	10 &	8 &	20 &	41 &	60 &	-19 &	38\\
15 &	Aston Villa &	38 &	10 &	8 &	20 &	39 &	61 &	-22 &	38\\
16 &	Hull City &	38 &	10 &	7 &	21 &	38 &	53 &	-15 &	37\\
17 &	West Bromwich Albion &	38 &	7 &	15 &	16 &	43 &	59 &	-16 &	36\\
\hline
18 &	Norwich City &	38 &	8 &	9 &	21 &	28 &	62 &	-34 &	33\\
19 &	Fulham &	38 &	9 &	5 &	24 &	40 &	85 &	-45 &	32\\
20 &	Cardiff City &	38 &	7 &	9 &	22 &	32 &	74 &	-42 &	30\\
\hline
\end{tabular}
\end{table}

\begin{figure}
      \centering
      \includegraphics[scale=0.7]{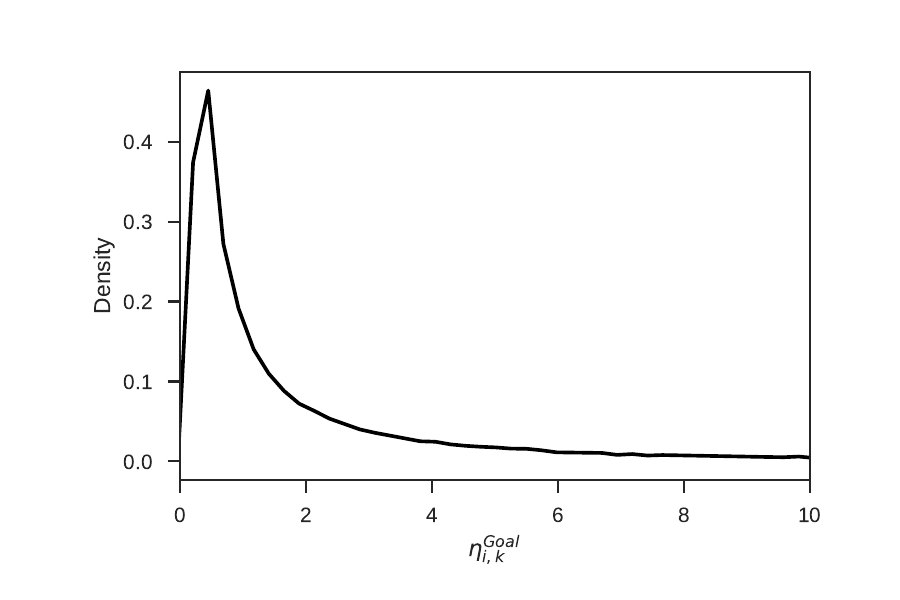}
      \caption{Monte Carlo simulation of the prior for $\eta_{i,k}^{\textrm{Goal}}$ using 100K draws of $\Delta_i^{\textrm{Goal}}$ from \eqref{ab-prior}.} \label{priorgoal}
\end{figure}

For a pair of interacting event types we fit the model defined by 
\eqref{posmod}--\eqref{llike}, by maximising \eqref{model_elbo} where 
$q(\cdot)$ follows \eqref{qDelta_single}. This model setup has 2182 
parameters 
$[544\times(\mu_{\Delta_i^{e_1}},\mu_{\Delta_i^{e_2}},\sigma_{\Delta_i^{e_1}},\sigma_{\Delta_i^{e_2}}) + \lambda_1^{e_1} + \lambda_2^{e_1} + \gamma^{\,e_1} + \lambda_1^{e_2} +  \lambda_2^{e_2} + \gamma^{\,e_2}]$ 
governing any two interacting event types. We take a  
reasonably uninformative prior 
\begin{equation}\label{ab-prior}
\pi\left(\Delta_i^{e}\right)\sim N\left(-2,2^2\right),
\end{equation}
where $-2$ represents the ability of an average player. We found little 
difference in results for alternative priors. We begin by considering 
occurrences of Goal and GoalStop. GoalStop is an event type of our own 
creation (in conjunction with expert football analysts at Stratagem Technologies), made up of 
many other event types (BallRecovery, Challenge, Claim, Error, 
Interception, KeeperPickup, Punch, Save, Smother, Tackle), with 
BallRecovery being the event type where a player collects the ball 
after it has gone loose. GoalStop aims to represent all the things a 
team can do to stop the other team from scoring a goal. A Monte Carlo 
simulation of the prior for $\eta_{i,k}^{\textrm{Goal}}$ using 100K 
draws of $\Delta_i^{\textrm{Goal}}$ from~\eqref{ab-prior} is shown in 
Figure~\ref{priorgoal}, where most players are viewed to score 0 or 1 
goal (as expected).

\begin{figure}[t]
\begin{minipage}[b]{0.48\linewidth}
        \centering
        \includegraphics[scale=0.55]{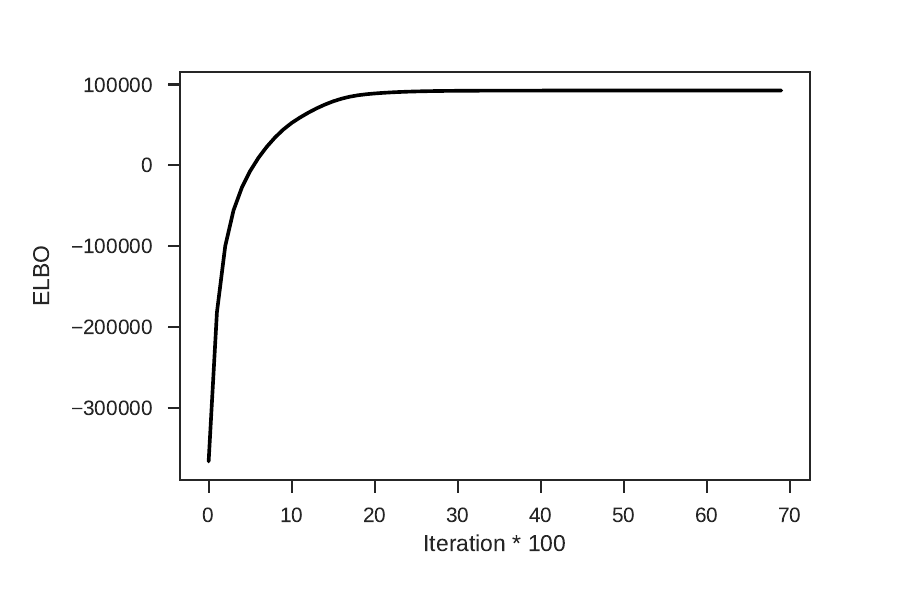}
\end{minipage} 
\begin{minipage}[b]{0.48\linewidth}
        \centering
        \includegraphics[scale=0.55]{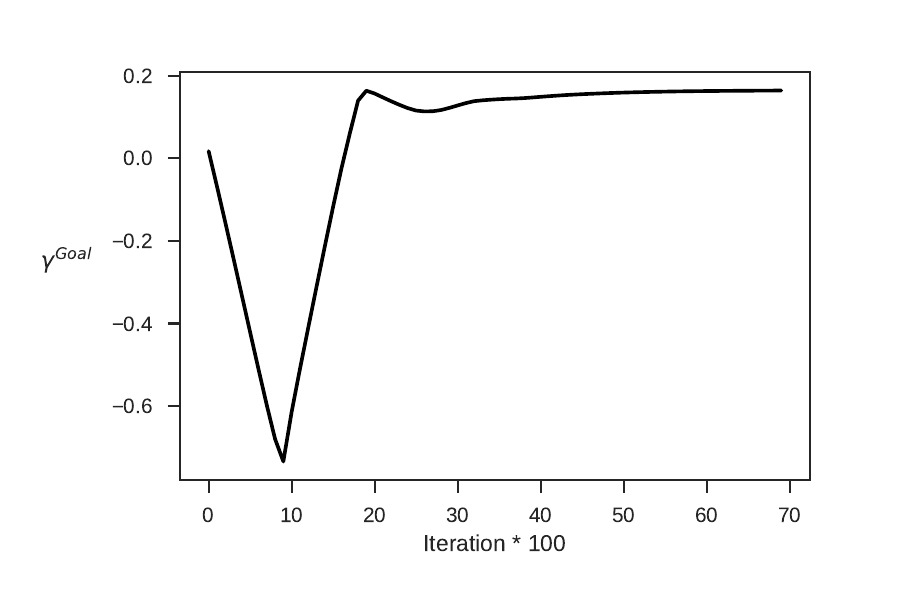}
\end{minipage} \\
\begin{minipage}[b]{0.48\linewidth}
        \centering
        \includegraphics[scale=0.55]{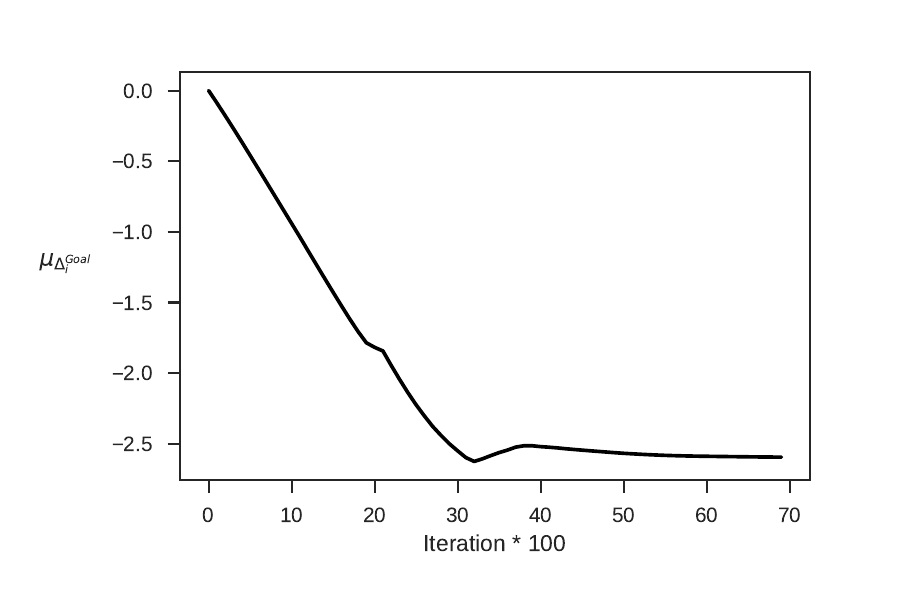}
\end{minipage} 
\begin{minipage}[b]{0.48\linewidth}
        \centering
        \includegraphics[scale=0.55]{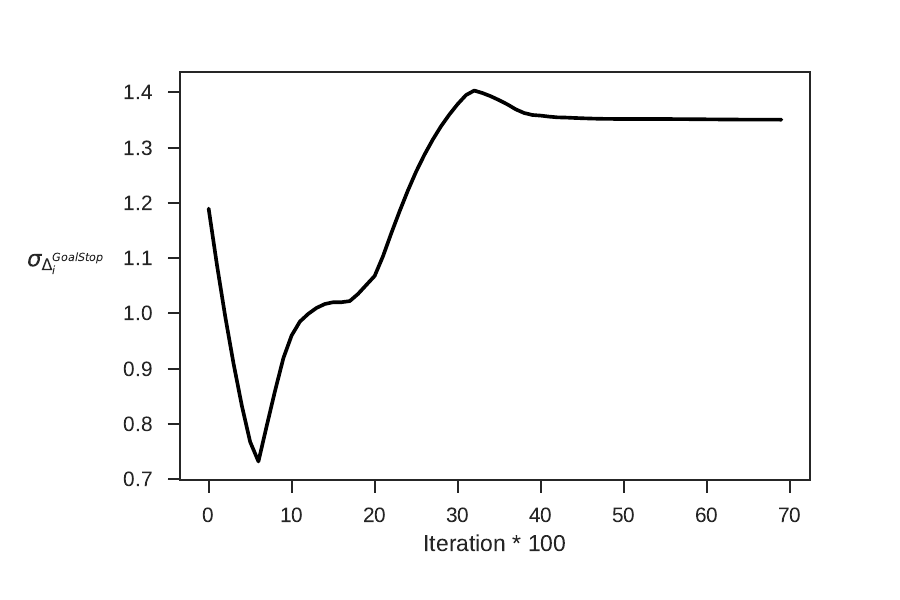}
\end{minipage} 
      \caption{Trace plots for the ELBO and a selection of the model parameters outputted every 100 iterations.} \label{paramfig}
\end{figure} 

We ran the model for 7000 iterations to achieve convergence. Trace 
plots of the ELBO and a selection of model parameters are shown in 
Figure~\ref{paramfig}. It is clear that convergence has been achieved 
(measured via the ELBO). For completeness the values of the fixed parameters 
($\psi$) under the model are given in Table~\ref{tableparam}, where 
all respective parameters appear to be on the same scale. We observe small 
differences in the parameters dictating the amount of impact both a 
player's own team, and the opposing team has on occurrences of an 
event type. There are more noticeable differences in the home effects 
of each event type, with the home effect for Goal being much larger 
than that of GoalStop. This is in line with other research around the 
goals scored in a match, where a clear home effect is acknowledged; 
see \cite{dixon_1997}, \cite{karlis_2003} and \cite{baio_2010} 
for further discussion of this home effect. The home effect for 
GoalStop is closer to zero, suggesting the number of attempts a team 
makes to stop a goal is similar whether they are playing at home or 
away.

\begin{figure}[t]
\begin{minipage}[b]{0.48\linewidth}
        \centering
        \qquad\qquad Goal\vspace{0.01cm}
        \includegraphics[scale=0.55]{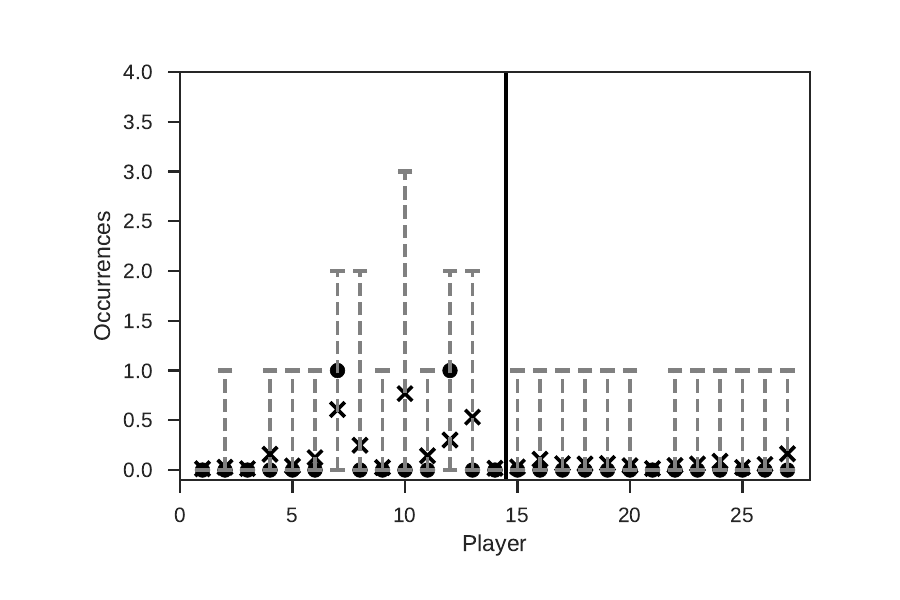}
\end{minipage} 
\begin{minipage}[b]{0.48\linewidth}
        \centering
        \qquad\qquad GoalStop\vspace{0.01cm}
        \includegraphics[scale=0.55]{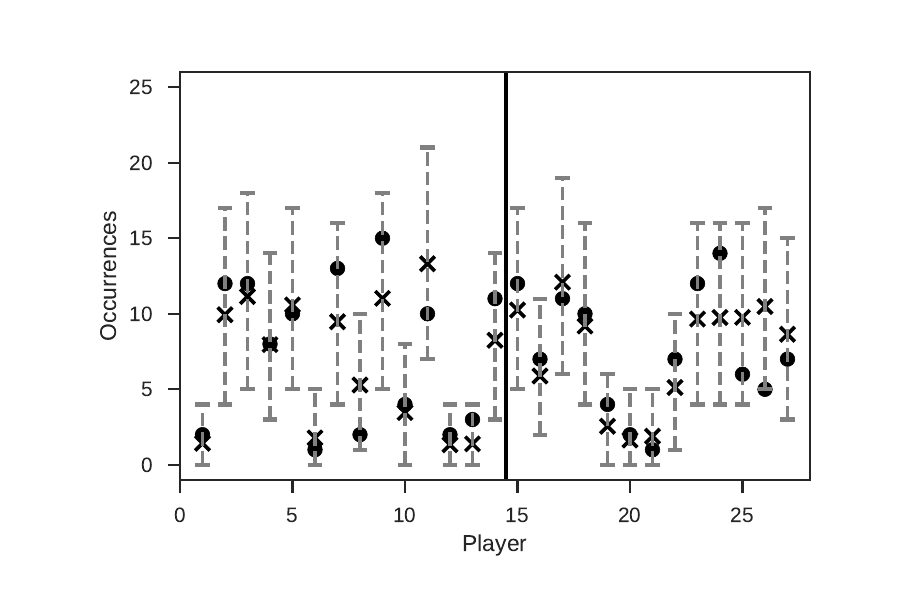}
\end{minipage} \\
\begin{minipage}[b]{0.48\linewidth}
        \centering
        \includegraphics[scale=0.55]{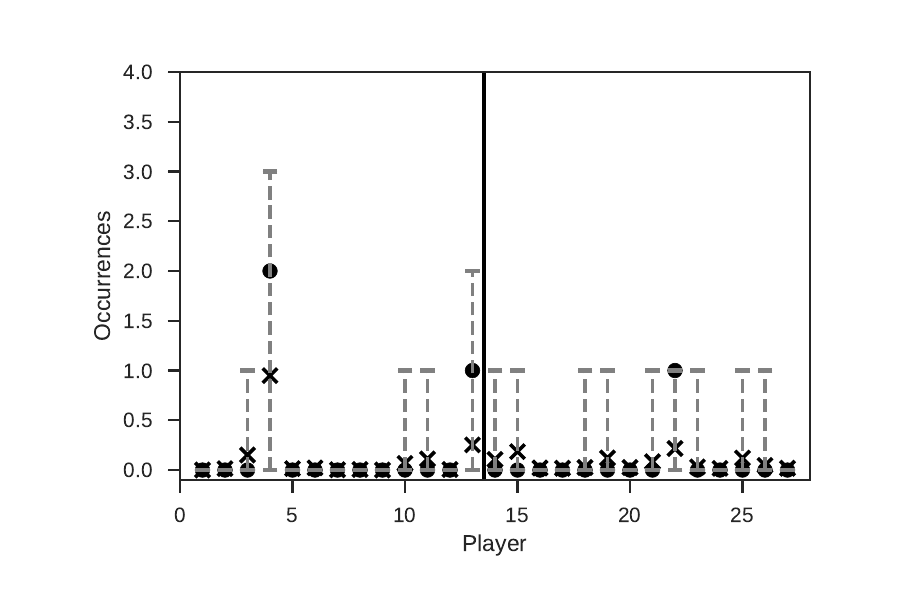}
\end{minipage} 
\begin{minipage}[b]{0.48\linewidth}
        \centering
        \includegraphics[scale=0.55]{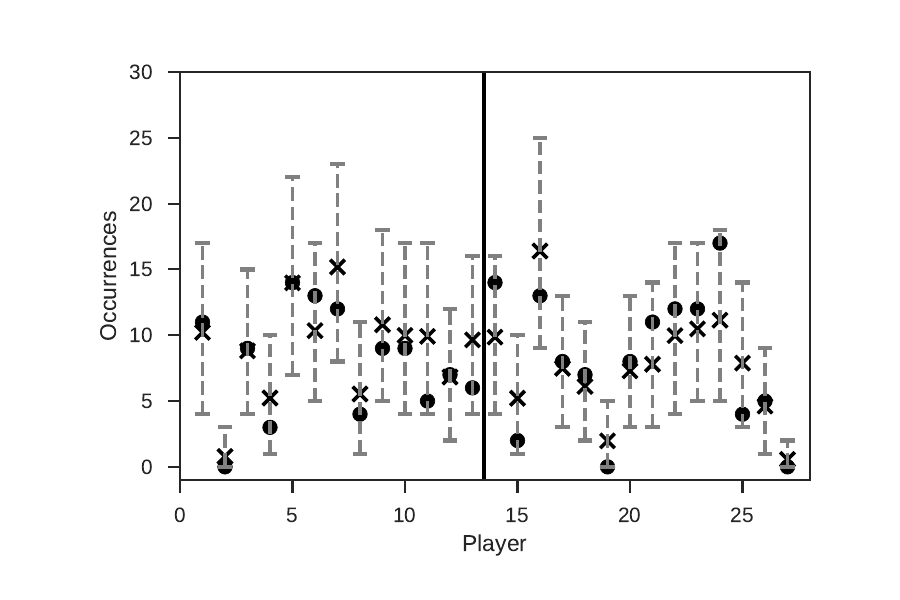}
\end{minipage} 
      \caption{Within sample predictive distributions for the number of goals/goal-stops in 2 randomly selected matches. \emph{Cross} model combinations of $\eta_{i,k}^{e}$, \emph{circle}~observed counts, \emph{dashed bars} 95\% prediction interval for each~$\eta_{i,k}^{e}$. The solid line separates the players from the two teams. The teams involved in the matches are Cardiff City, Hull City, Liverpool and Manchester City.} \label{fitfig}
\end{figure}

Figure~\ref{fitfig} shows the $\eta_{i,k}^{e}$ \eqref{eta} we obtain when the 
model parameters are combined for 2 randomly selected matches, where we 
set $\Delta_i^{e}$ to be $\mu_{\Delta_i^{e}}$. We plot these 
against the observed counts and include the 95\% prediction intervals 
for each $\eta_{i,k}^{e}$ to add further clarity. The solid line on 
each plot separates the players from the two opposing teams. A large 
number of the model $\eta$s are close to the observed counts 
(especially for GoalStop), and nearly all observed values fall within 
the 95\% prediction intervals. The 
number of goal-stops across teams is not particularly variable, 
however there is some evidence of player variability (although this is 
somewhat clouded by the fact that substitutes are not specifically 
marked in the figure, as we would expect them to register lower counts by virtue of 
less playing time).

\begin{figure}[t]
      \centering
      \includegraphics[scale=0.4]{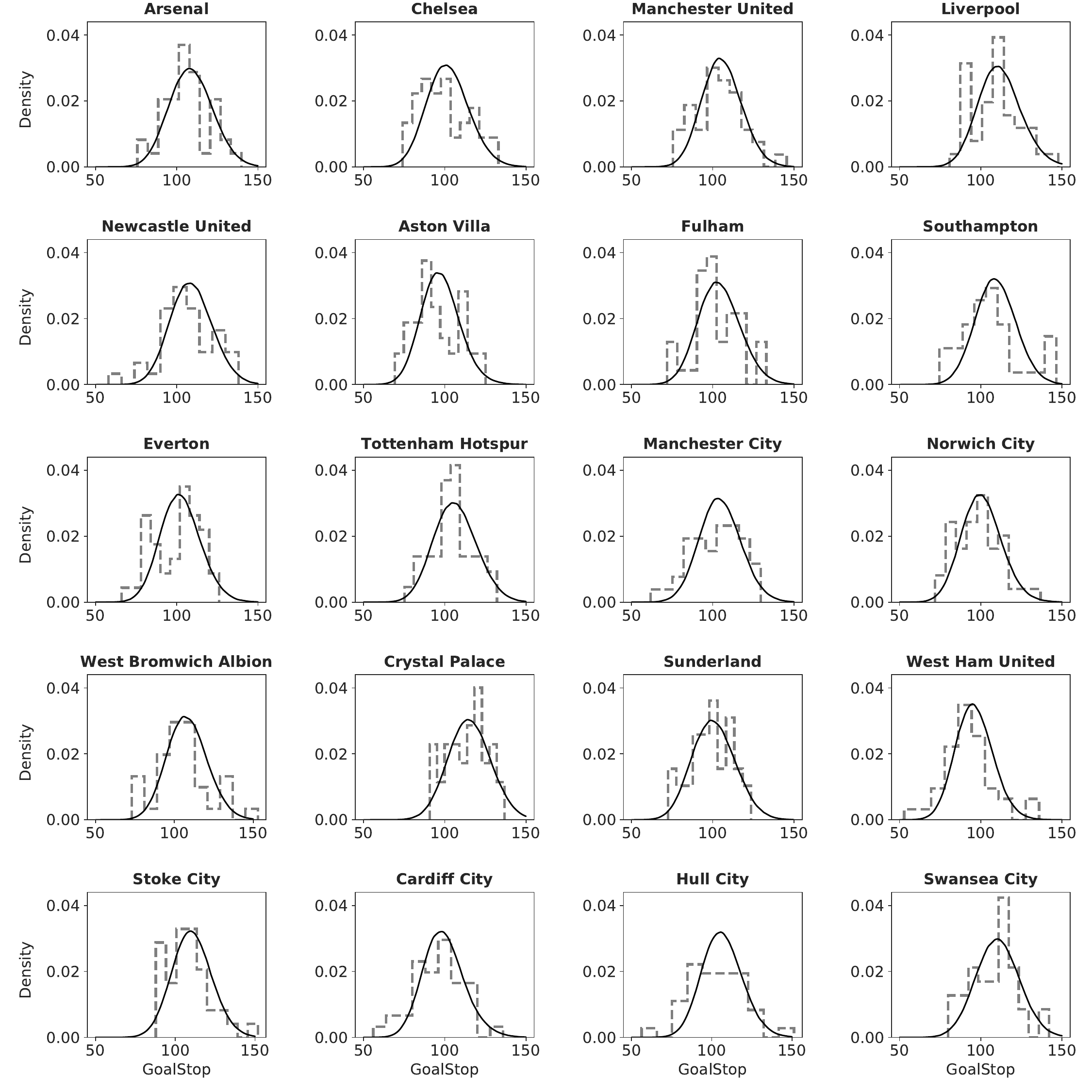}
      \caption{Within sample predictive distributions for the total number of goal-stops in a game for each team in the 2013/2014 English Premier League \emph{solid}. The observed totals from the 38 games is given by the histograms \emph{dashed}.}  \label{goalstopdensfig}
\end{figure}

We sample the marginal posterior variational densities, $q(\Delta_i^{e})$, 
10K times (constructing the corresponding $\eta_{i,k}^{e}$) and simulate 
from the relevant Poisson distributions (with mean 
$\eta_{i,k}^{e}\tau_{i,k}$). This gives a Monte Carlo simulation of 
each player's number of goals and goal-stops for each game in the 
2013/2014 English Premier League. Summing over the players who played 
in a given match gives a within sample prediction of the 
total number of goals/goal-stops for each team in every game. We 
present the predictive distributions of these totals for GoalStop in 
Figure~\ref{goalstopdensfig}, where for reference we also include 
histograms for each team's total number of goal-stops in each 
game constructed from the touch-by-touch data. The model is clearly 
capturing the patterns between differing teams (and the patterns 
observable within the data). We note, slightly surprisingly, 
that there appears to be no connection between the occurrences of 
GoalStop and the goals a team concedes, with both Chelsea and Norwich 
City having similar predictive distributions despite conceding a 
vastly different number of goals, 27 and 62 respectively. There is 
some suggestion that such observations may be used to determine a 
team's style of play, for example, whether they are a \emph{passing} 
team or follow the \emph{long ball} philosophy. We leave such 
questions for future investigation given the setup we derive here. We 
can conclude, nevertheless, that the model is capturing the trends 
observed in the touch-by-touch data well.

\begin{figure}
\begin{minipage}[b]{0.48\linewidth}
        \centering
        \qquad\quad Sturridge\vspace{0.01cm}
        \includegraphics[scale=0.55]{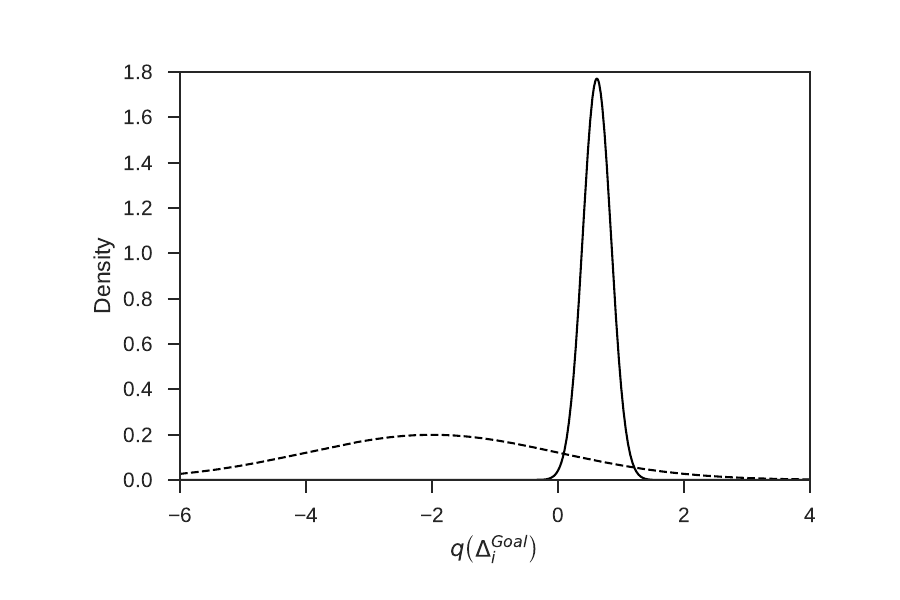}
\end{minipage} 
\begin{minipage}[b]{0.48\linewidth}
        \centering
        \qquad\quad Reed\vspace{0.01cm}
        \includegraphics[scale=0.55]{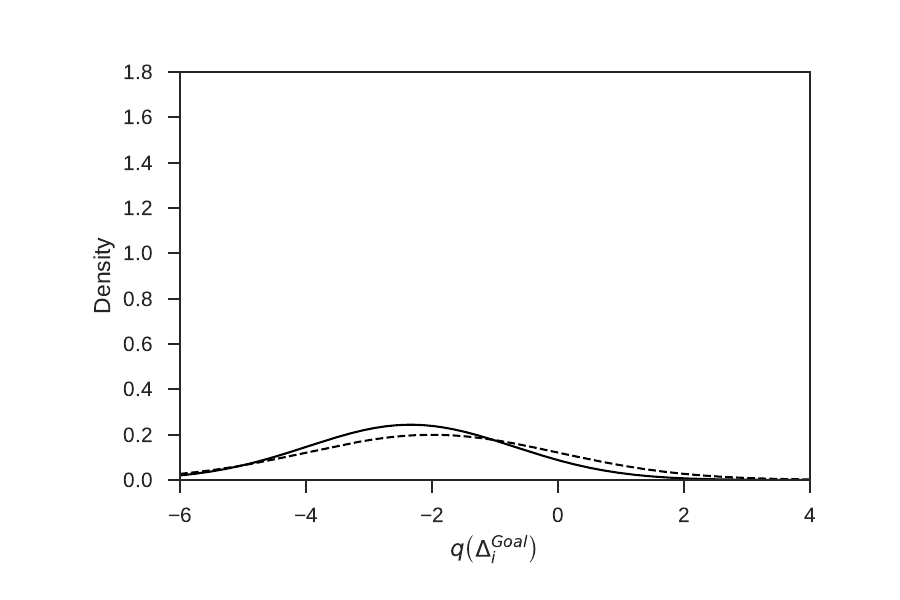}
\end{minipage} \\
      \caption{Marginal posterior variational densities of Goal for 2 players in the 2013/2014 English Premier League. \emph{Dashed} prior, \emph{solid} posterior.} \label{playerfig}
\end{figure}

Marginal posterior variational densities of Goal, 
$q(\Delta_i^{\textrm{Goal}})$, for two players are presented in 
Figure~\ref{playerfig}, where $q(\Delta_i^{\textrm{Goal}})$ takes 
the form of \eqref{qDelta_single} and the prior is \eqref{ab-prior}. 
The two players shown are Daniel Sturridge and Harrison Reed. 
Sturridge played 29 times over the season totalling 2414 minutes of 
match time, scoring 21 goals, whereas Reed played 4 times totalling 
23 minutes, scoring zero goals. These attributes are clearly captured 
by the posteriors; the greater number of observations for Sturridge 
leading to a posterior with a much smaller variance. The high number 
of goals scored by Sturridge leads to him having a higher value of 
$\mu_{\Delta_i^{\textrm{Goal}}}$ (with reasonable certainty), whilst 
the lack of both goals and playing time leads to a posterior for Reed 
which resembles the prior.

\begin{table}[h] 
\caption{Values of the fixed parameters ($\psi$) for interacting event types Goal and GoalStop.} \label{tableparam}
\centering
\begin{tabular}{l|ccc}
\hline
& \multicolumn{3}{c}{Fixed parameter}\\
Event type ($e$) & $\lambda_1^e$ & $\lambda_2^e$ & $\gamma^e$  \\[0.3ex]
\hline \\[-1.9ex]
Goal & $2.907\times 10^{-8}$ & 0.041 & 0.165 \\[0.3ex]
GoalStop & $1.621\times 10^{-7}$ & 0.009 & 0.003 \\
\hline
\end{tabular}
\end{table}

The model is capturing differences between players abilities, 
as evidenced by the posteriors of Figure~\ref{playerfig}. Thus, the 
natural question to ask is whether these differences are sensible, and,
if we were to order the players by their inferred ability, would this 
ordering agree with (a debatable) reality. Hence, we construct the 
marginal posterior variational densities for all players and rank them 
according to the 2.5\% quantile of these densities. A top 10 for Goal 
is presented in Table~\ref{tablegoal}, with a ranking for GoalStop 
given in Table~\ref{tablegoalstop}. We present top 10 lists for other 
event types in Appendix~\ref{app_top10}. The ranking shown in 
Table~\ref{tablegoal} appears sensible and comprises of those players 
who were the main goal scorers (strikers) for the best teams, and the 
players who scored nearly all the goals a lesser team scored over the 
season. The ranking is very close to that obtained by ranking players 
on the total number of goals scored over the season (although there is 
some debate in the football community as to whether this is a sensible 
way of ranking, with some suggesting a ranking based on a per 90 minute 
statistic, however this can be distorted by those with very little 
playing time, see Chapter~3 of \cite{anderson_2013} or \cite{AGR_2016} for further discussion). 
The questionable deviations from this ranking are Aguero 
(ranked third) and van Persie (ranked seventh). Both these players have 
less playing time over the season compared to their competitors, and 
thus, the model highlights them as better goal scorers, given the time 
available to them, than other players based on total goals scored. 
Expert football analysts agreed with this view when we showed them 
these rankings. Suarez has an inferred ability much greater than any other player, 
which is evidenced by the 31 goals he scored (10 more than any other 
player). At points the difference between successive ranks is small, 
suggesting some players are harder to distinguish between. Finally, we 
note that the standard deviations for all players in the top 10 are 
roughly the same, meaning we have similar confidence in the ability 
of any of these players.

\begin{table}[t] 
\caption{Top 10 goal scorers in the 2013/2014 English Premier League based on the 2.5\% quantile of the marginal posterior variational density for each player, $q(\Delta_i^{\textrm{Goal}})$.} \label{tablegoal}
\centering
\footnotesize
\begin{tabular}{cllccccccc}
\hline
\multicolumn{10}{c}{Goal - top 10}\\
\hline
\multirow{2}{*}{Rank} & \multirow{2}{*}{Player} &  \multirow{2}{*}{Team} & 2.5\% & \multirow{2}{*}{Mean} & Standard & \multirow{2}{*}{Observed} & Observed & Rank & Time\\
 &  &  & quantile & & deviation &  & rank & difference & played \\
\hline
1 &  Suarez 	& Liverpool	& 0.508 & 0.869 &	0.184	& 31	& 1 & 0  & 3185\\ 
2 &  Sturridge & Liverpool	& 0.176 & 0.617 &	0.225  	& 21	& 2 & 0  & 2414\\
3 &  Aguero 	& Manchester City	& 0.147 & 0.636 &	0.250	& 17	& 4 & +1  & 1616\\
4 & Y. Toure 	& Manchester City	& -0.043 & 0.395 &	0.224	& 20	& 3 & -1  & 3113\\
5 & Rooney 	& Manchester United	& -0.056 & 0.421 &	0.243	& 17	& 5 & 0  & 2625\\
6 & Dzeko 	&	Manchester City & -0.065 & 0.424 &	0.249 		& 16	& 8 & +2  & 2128\\
7 & van Persie & Manchester United	& -0.136 & 0.430 &	0.289 	& 12	& 15 & +8  & 1690 \\
8 & Remy 	& Newcastle United	& -0.230 & 0.302 &	0.271 	& 14	& 11 & +3  & 2274\\
9 & Bony 	& Swansea City	& -0.257 & 0.238 &	0.252 	& 16	& 7 & -2  & 2644\\
10 & Rodriguez 	& Southampton	& -0.354 & 0.161 &	0.263 	& 15	& 10 	& 0  & 2758\\
\hline
\end{tabular}
\end{table}

The ranking of GoalStop (Table~\ref{tablegoalstop}) appears at first 
glance to be less sensible than that of Table~\ref{tablegoal}. It 
features 3 players with comparatively larger standard deviations, 
Kallstrom (rank 2), Lewis (rank~6) and Palacios (rank 7). Whilst these 
players did well with the little playing time afforded to them, it is 
somewhat presumptuous to postulate that they would maintain a similar 
level of ability given more game time, leading to their ranking slipping. 
Ideally we would provide several tables for each 
event type, filtering players by the amount of uncertainty surrounding 
them, although such an approach would be unwieldy given the large 
number of players in the dataset. Moreover, fully factorised mean-field 
approximations are known to underestimate the uncertainty of the 
posterior \citep{bishop_2006}. Although comparative uncertainty between 
players is easier to gauge, it is less clear how to quantify how much 
bias is being added to the variances of each latent variable 
individually. In a future work this could be mitigated by adopting a 
variational approximation that accounts for some correlations of the 
latent variables, or by correcting for the variances using the approach 
of \cite{giordano_2018}. However, the rest of the list appears sensible and is 
made up mainly of defensive midfielders (whose main role it is to 
disrupt the oppositions play); only Mannone and Ruddy are goalkeepers 
(discounting the 3 players with large standard deviations). This 
suggests, anecdotally given our definition of GoalStop, that to 
stop a goal it is more prudent to invest in a better 
defensive midfielder than it is a goalkeeper, presuming you can not 
just buy the best player in each position. Here, the differences 
between successive ranks are much smaller than in Table~\ref{tablegoal}, 
implying it is harder to distinguish between player ability to perform 
goal-stops than it is the ability to score goals.

\begin{table}
\caption{Top 10 goal-stoppers in the 2013/2014 English Premier League based on the 2.5\% quantile of the marginal posterior variational density for each player, $q(\Delta_i^{\textrm{GoalStop}})$.} \label{tablegoalstop}
\centering
\footnotesize
\begin{tabular}{cllccccccc}
\hline
\multicolumn{10}{c}{GoalStop - top 10}\\
\hline
\multirow{2}{*}{Rank} & \multirow{2}{*}{Player} &  \multirow{2}{*}{Team} & 2.5\% & \multirow{2}{*}{Mean} & Standard & \multirow{2}{*}{Observed} & Observed & Rank & Time\\
 &  &  & quantile & & deviation &  & rank & difference & played \\
\hline
1 & Mulumbu & West Bromwich Albion &	2.575 & 2.653 &	0.040  & 	631	  & 1 & 0 & 3319\\
2 & Kallstrom & Arsenal &	2.553   & 2.900 &	0.177  & 	33	    & 405 & +403 & 144 \\
3 & Mannone & Sunderland &	2.528 	  & 2.615 &	0.044  & 	508	    & 12 & +9 & 2767  \\
4 & Yacob  & West Bromwich Albion &	2.510 	& 2.614 &	0.053  &	359	  & 43 & +39 & 1979 \\
5 & Tiote  & Newcastle United &	2.474   & 2.560 &	0.044  & 	517	    & 8 & +3 & 2988 \\
6 & Lewis & Cardiff City	 &	2.446 	  & 2.863 &	0.213  & 	23	    & 436 & +430 & 98 \\
7 & Palacios & Stoke City &	2.441 &	2.638 &	0.101  &	100	    & 286 & +279 & 585 \\
8 & Jedinak & Crystal Palace &	2.420 	  & 2.500 &	0.041  & 	603	    & 2 & -6 &  3651 \\
9 & Ruddy &	Norwich City &	2.411 	  & 2.491 &	0.041  & 	600	    & 3 & -6 &  3679\\
10 & Arteta &	Arsenal & 	2.409 &	2.503 &	0.048  & 	431	  & 21 & +11 & 2615 \\
\hline
\end{tabular}
\end{table}

\newpage
Overall, the model provides a good fit to the data and suggests 
a reasonable prowess to determine a player's ability in a specific 
event type, with marginal posterior variational densities providing a 
good visual comparison between different players abilities (and the 
confidence surrounding that ability).

\subsubsection*{Sensitivity analysis for the contribution of {\boldmath$\tau_{i, k}$}} 

To address questions remaining about the inclusion of $\tau_{i,k}$ in~\eqref{eta}, we investigate the validity of this decision by considering the predictive 
performance of ~\eqref{eta} versus a simplified version of~\eqref{eta} 
where $\tau_{i,k}$ is suppressed. We test both versions of the model 
on the scenario described in Section~\ref{ability}. Calculating 
out-of-sample prediction biases for the events Goal, GoalStop and Shots 
we found that the model described in this paper gave 0.201, 4.374 and 0.807 
respectively, with 0.237, 5.130, 0.807 observed for the model without 
$\tau_{i,k}$.
We see an improved prediction bias for Goal and GoalStop when the model 
includes $\tau_{i,k}$ with the same result seen under both models for 
Shots. As our ultimate aim within this paper lies with prediction, we 
therefore favour the model including time described by 
\eqref{posmod} and~\eqref{eta}. The inclusion of time is not an obvious 
one, and whilst it improves prediction, that on it's own is not reason 
enough. We believe by acting as a proxy for how much a player 
interacts with their team in an event type and how put off they would 
be by the opposition, it may also capture some of the more nuanced 
aspects of a football game, for example, a player's importance in a 
manager's formation. Capturing the many nuances affecting a player in 
a football game would be an interesting area for future research, 
although given the near endless possibilities up for inclusion, the form 
of the model used in this paper may be restrictive. In the next section we look to 
utilise the player abilities explored in Section~\ref{ability} in the prediction of goals in a football 
match.

    \subsection{Prediction} \label{prediction}
    
A key betting market stemming from the rise of online betting is the 
over/under market \citep{betfair_2017, bethq_2017, Sportingindex_2017}, 
where people bet on whether over or under 2.5 goals 
will be scored in a match. Here we attempt to 
predict whether 2.5 goals will be scored or not in a given game.  
To predict the goals scored in a game which takes place in the 
future we first fit the model on all the past available to us. We 
use a whole season of data to train the 
model, before predicting the following season in incremental blocks. 
Here, we use the entirety of the 2013/2014 English Premier League 
season (380 games) to train the model, before attempting to predict 
the goals scored in each match of the 2014/2015 English Premier League 
season. The reason why we use the whole of the 2013/2014 season to 
train the model, and only predict on the 2014/2015 season, is that the 
odds data available to us only covers the 2014/2015 season. 
We introduce the games (on which we predict) in blocks of 
size 80, with a final block of 60 games to total 380 (the number of 
games over a season). In each case we use all of the available past 
to fit the model, that is, in predicting the second block of 80 games in 
the 2014/2015 season, we use all of the 2013/2014 season and the first 
block of the 2014/2015 season to fit the model. Figure~\ref{pic-window} 
shows a graphical representation of this approach.

\begin{figure}
\centering
\begin{tikzpicture}[scale=5,>=latex]
     \draw[black!20!,-]
         (0,0) -- (0,0.5);
     \draw[black!20!,-]
         (1,0) -- (1,0.5);         

     \node[draw,circle,fill=black,inner sep=0.5mm, label=above:{13/14}]
         (xo0) at (0,0.5) {};
     \node[draw,circle,fill=black,inner sep=0.5mm, label=above:{14/15}]
         (xo1) at (1,0.5) {};
         
     \draw[black]
         ((0,0.35) rectangle (1,0.45);
     \draw[black!70!,densely dotted]
         ((1,0.35) rectangle (1.25,0.45);
           
     \draw[black]
         ((0,0.2) rectangle (1.25,0.3);
     \draw[black!70!,densely dotted]
         ((1.25,0.2) rectangle (1.5,0.3);  
     
     \draw[black]
         ((0,0.05) rectangle (1.5,0.15);
     \draw[black!70!,densely dotted]
         ((1.5,0.05) rectangle (1.75,0.15);    
           
     \node at (1.895, -0.01) {\vdots\;etcetera};    
        
\end{tikzpicture}
\caption{Illustration of the approach to prediction. The model is fit on all past data before predictions are made for a future block of games. \emph{Solid} fit, \emph{dashed} predict.} \label{pic-window}
\end{figure} 

For the extension to the model of \cite{baio_2010} (which we consider 
to be the baseline model), we include the latent player abilities for 
the event types Goal, Shots and ChainEvents, with their counterparts 
being GoalStop, ShotStop and AntiPass respectively. Goal and GoalStop 
are as defined in Section~\ref{ability}, whilst Shots and ShotStop 
have homogeneous roots to Goal and GoalStop, that being the ability to 
shoot or to stop a shot. ChainEvents represents how prevalent a player 
is in the lead up to a good attacking chance, with AntiPass being a 
player's ability to stop the other team from passing the ball. We 
refer the reader to Appendix~\ref{app_top10} for the more technical 
definitions of these event types. Explicitly \eqref{fdeltah} and 
\eqref{fdeltaa} are given by 
\begin{align}
f\left(\Delta\right)_h &= \sum_{i\in I_k^{T_k^H}}\left(\Delta_{i}^{\textrm{Goal}} + \Delta_{i}^{\textrm{Shots}} + \Delta_{i}^{\textrm{ChainEvents}}\right) \nonumber\\
&\qquad\qquad- \sum_{i\in I_k^{T_k^A}}\left(\Delta_{i}^{\textrm{GoalStop}} + \Delta_{i}^{\textrm{ShotStop}} + \Delta_{i}^{\textrm{AntiPass}}\right) \label{fdeltah-pred} 
\shortintertext{and} 
f\left(\Delta\right)_a &= \sum_{i\in I_k^{T_k^A}}\left(\Delta_{i}^{\textrm{Goal}} + \Delta_{i}^{\textrm{Shots}} + \Delta_{i}^{\textrm{ChainEvents}}\right) \nonumber\\
&\qquad\qquad- \sum_{i\in I_k^{T_k^H}}\left(\Delta_{i}^{\textrm{GoalStop}} + \Delta_{i}^{\textrm{ShotStop}} + \Delta_{i}^{\textrm{AntiPass}}\right), \label{fdeltaa-pred}
\end{align}
where $I_k^j$ is the initial eleven players who start game $k$ for 
team $j$. 
We also considered including a player's ability to pass, but found 
this led to no increase in predictive power (and in some instances 
diminished it). We found little difference when setting $\Delta_i^{e}$ 
to be either $\mu_{\Delta_i^{e}}$ or the 2.5\% quantile of 
$q(\Delta_i^{e})$ in \eqref{fdeltah-pred} and \eqref{fdeltaa-pred}, 
and so here report results for the mean ($\mu_{\Delta_i^{e}}$).    

Both the models were fit using \verb+PyStan+ and were run long enough 
to yield a sample of approximately 10K independent posterior draws, 
after an initial burn in period. The teams which feature in the data 
are those in Table~\ref{finaltable}, with the addition of Burnley, 
Leicester City and Queens Park Rangers who replaced the relegated 
teams of Cardiff City, Fulham and Norwich City for the 2014/2015 
season. The setup outlined at the beginning of this section allows us to view the 
evolution of a team's attack/defence parameter or a player's latent 
ability through time after different fitting blocks. We denote block 
0 to be all the games in the 2013/2014 English Premier League 
season, block 1 to be block 0 plus the first 80 games of the 
2014/2015 season, block 2 to be block 1 plus the next 80 games, 
block 3 to include the next 80 games, with block 4 including the 
next 80 games again. As an indication, for block 1, the mean predictive  
log-likelihood for the model including the latent player abilities is 
$-163.106$ with the baseline model having a mean predictive log-likelihood 
of $-159.578$ (this is the log-likelihood of the second 80 games of the 
2014/2015 season having fit the models using data in block 1). 
We observe similar results across all other blocks. However, 
we believe the main judgement of the models within the context of this 
paper will be the quality of the predictions they produce and not solely 
the log-likelihood. 

\begin{figure}[t]
\begin{minipage}[b]{0.48\linewidth}
        \centering
        \qquad\, Baseline\vspace{0.01cm}
        \includegraphics[scale=0.55]{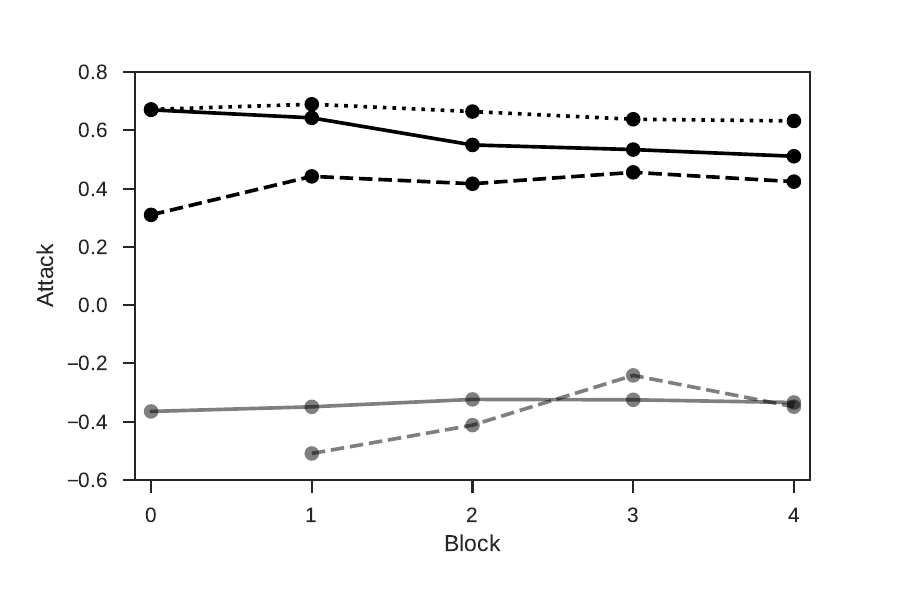}
\end{minipage} 
\begin{minipage}[b]{0.48\linewidth}
        \centering
        \qquad\, Including latent player abilities\vspace{0.01cm}
        \includegraphics[scale=0.55]{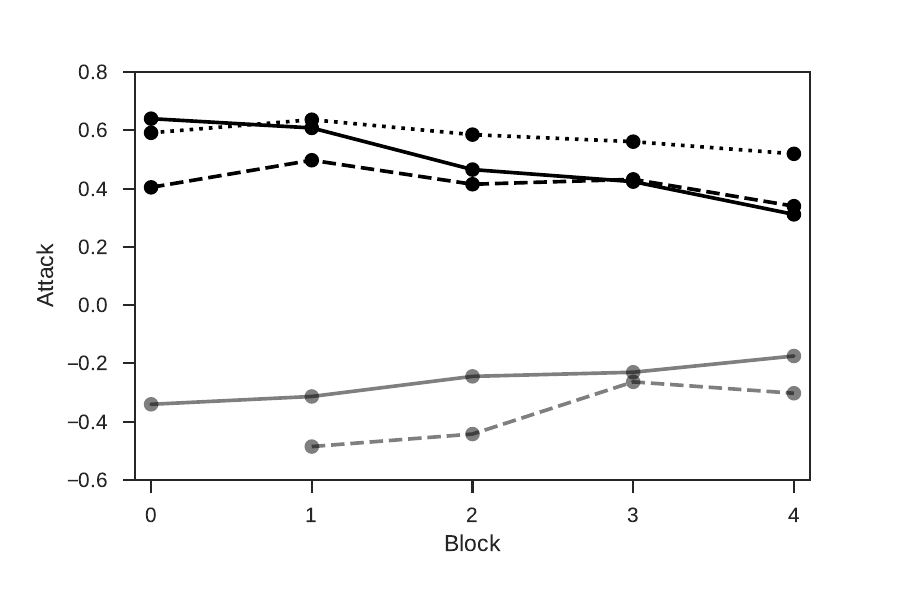}
\end{minipage} \\
\begin{minipage}[b]{0.48\linewidth}
        \centering
        \includegraphics[scale=0.55]{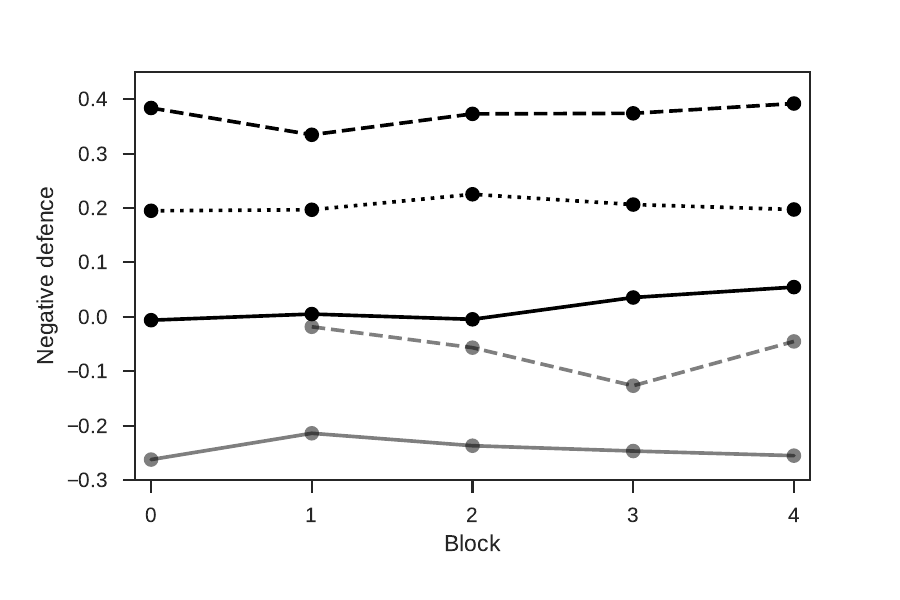}
\end{minipage} 
\begin{minipage}[b]{0.48\linewidth}
        \centering
        \includegraphics[scale=0.55]{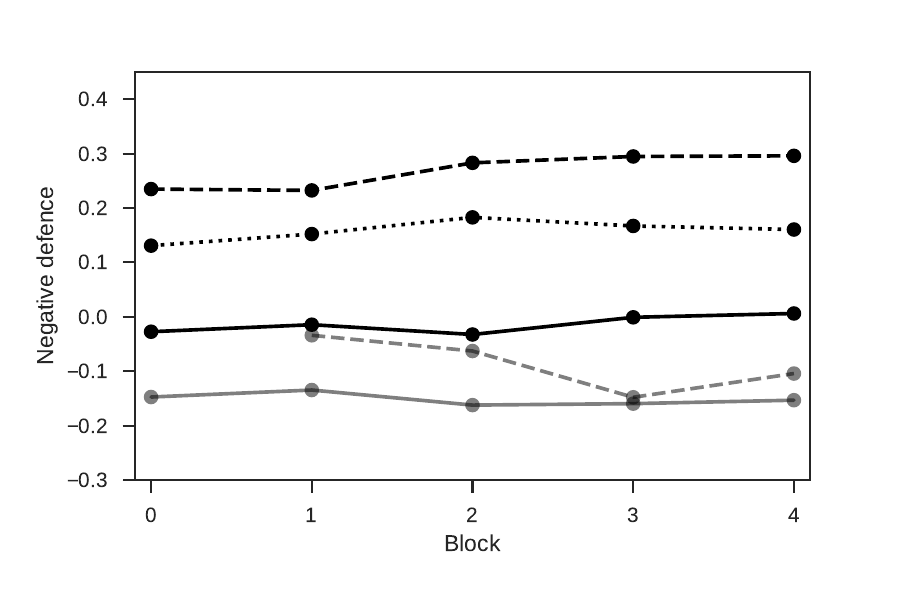}
\end{minipage}
      \caption{Attack and defence parameters through time under the baseline model and the model including the latent player abilities for selected teams. \emph{Top row} attack, \emph{bottom row} negative defence. \hbox{\emph{Black-solid} Liverpool}, \emph{black-dashed} Chelsea, \emph{black-dotted} Manchester City, \emph{grey-solid} Cardiff City, \hbox{\emph{grey-dashed} Burnley}.} \label{attdeffig}
\end{figure} 

The attack and defence parameters through time 
for both the baseline model and the model including the latent player 
abilities for selected teams are shown in Figure~\ref{attdeffig}, where 
we plot negative defence so that positive values indicate increased 
ability. Recall that these parameters for all teams must sum to zero. 
We see similar, but not identical patterns under both models. The 
model including latent player abilities reduces the variance of the 
attack and defence parameters compared to the baseline model, 
suggesting the inclusion of the $\Delta$s accounts for some of a 
team's attack and defensive ability. Manchester City and Chelsea follow 
similar patterns under both models, with Chelsea clearly having the 
best defence parameter. Including the $\Delta$s impacts Liverpool's 
attacking ability, where the removal of Suarez (Liverpool's best 
attacking player who transferred to Barcelona between the 2013/2014 
and 2014/2015 seasons) clearly reduces Liverpool's attacking threat at 
a more drastic rate than the baseline model. This is in line with 
reality, where Liverpool only scored 52 goals over the 2014/2015 season 
compared to 101 goals the previous year. Cardiff City, who got 
relegated after the 2013/2014 season but feature in all blocks despite 
not being used for prediction (as games involving them can inform 
the attack and defence abilities of other teams), have relatively 
constant parameters under both models, accounting for the reduction in 
variance. Notable for Burnley is the peak/trough observed after 
block~3; this is due to the fact that Burnley were starting to look at 
the prospect of relegation and needed to start winning games, hence, 
they tried (and succeeded) to score more goals in order to win games, 
but found themselves more likely to concede goals in the process of 
doing so.

\begin{figure}[t]
      \centering
      \includegraphics[scale=0.7]{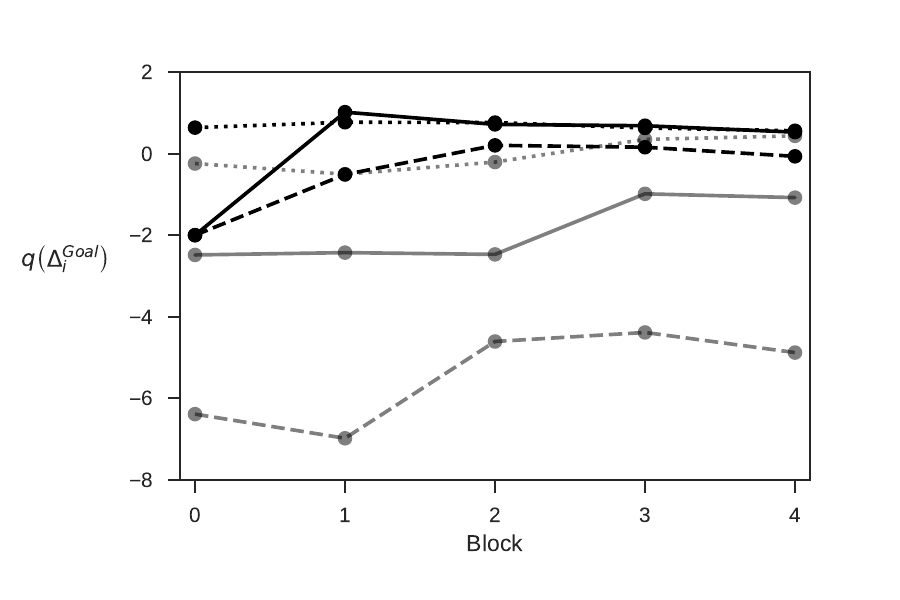}
      \caption{The mean of $q(\Delta_i^{\textrm{Goal}})$ through time for a selection of players. \hbox{\emph{Black-solid} Costa}, \hbox{\emph{black-dashed} A. Sanchez}, \emph{black-dotted} Aguero, \emph{grey-solid} Defoe, \emph{grey-dashed} G. Johnson, \hbox{\emph{grey-dotted} Kane}.} \label{goaltimefig}
\end{figure} 

\begin{figure}
      \centering
      \includegraphics[scale=0.7]{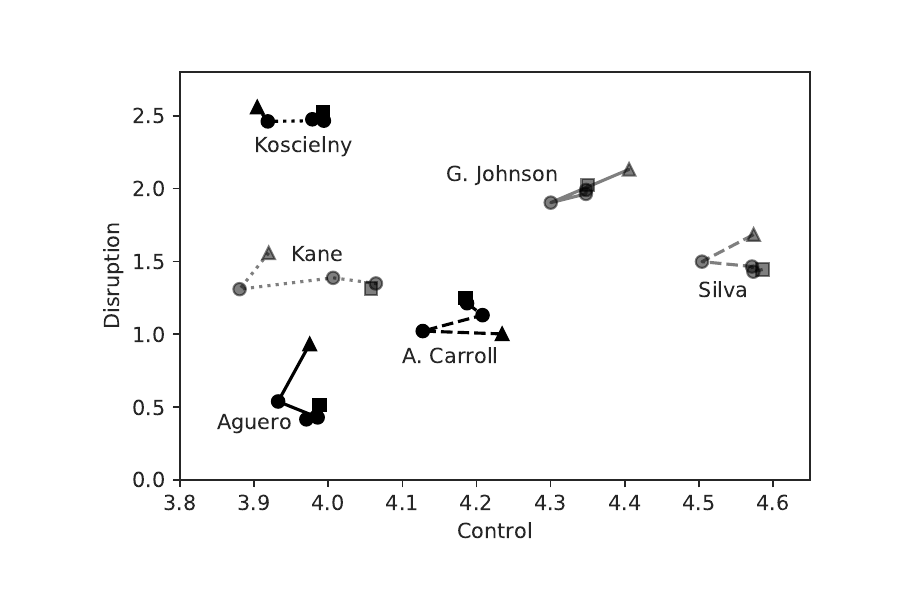}
      \caption{The mean of $q(\Delta_i^{\textrm{Control}})$ versus the mean of $q(\Delta_i^{\textrm{Disruption}})$ through time for a selection of players. \hbox{\emph{Triangle} block 0}, \emph{square} block 4.} \label{controltimefig} 
\end{figure} 

The mean of $q(\Delta_i^{\textrm{Goal}})$ through time for 
a selection of players are illustrated in Figure~\ref{goaltimefig}. We 
let this value represent a player's ability to score a goal. If a player 
has not featured previously in the data we represent their ability in the 
figure by the mean of the prior distribution ($-2$). We see that the model is 
quick to identify a given player's ability. To elucidate, Costa is 
immediately (after block~1) identified as one of the top goal scorers 
despite not featuring in the 2013/2014 season. The same can be said for 
Sanchez, who takes longer to establish his ability after a less 
impressive start to the season. Aguero was one of the best goal scorers 
across all the data and has a constant ability near the top. Kane had 
little playing time until block 2 where the model starts to increase 
his ability to score a goal. Defoe spent most of the 2013/2014 season 
on the bench before transferring to Toronto~FC; he returned to the 
English Premier League with Sunderland in January 2015, where he scored 
a number of goals, saving Sunderland from relegation. The model rightly 
acknowledges this and raises his ability as a goal scorer (a trait he 
is well known for). Given a player scores a small number of goals, 
relative to other event types, we include G. Johnson to show the effect 
of scoring a goal. Johnson scored 1 goal in the 2013/2014 and 2014/2015 
seasons (during block 2), and his ability rises by a large jump because 
of this; such jumps are not evident for players who score a reasonable 
number of goals (5+).

The mean of $q(\Delta_i^{\textrm{Control}})$ and the mean 
of $q(\Delta_i^{\textrm{Disruption}})$ are plotted against 
each other through time for a selection of players in 
Figure~\ref{controltimefig}. Control and Disruption comprise of the 
event types listed in Table~\ref{eventtype}. It is evident that for 
the majority of players their Control and Disruption abilities do not 
vary much through time (from block to block). This is perhaps 
unsurprising given we do not expect a player's ability to change 
dramatically from game to game. Those that vary the most are the 
players with fewer minutes played in the earlier blocks but have much 
more playing time as time progresses, for example, Kane (see 
Figure~\ref{controltimefig}). The figure does however show clear 
distinction between players, with defenders tending to occupy the top 
half of the graph and strikers the bottom. An interesting 
extension to this work would be to see whether a clustering analysis 
of these latent player abilities would reveal player positions, that is, 
central defender or wing-back for example.

\begin{figure}
\hspace{-1.2cm}
\begin{minipage}[b]{0.48\linewidth}
        \centering
        \qquad\qquad\qquad\quad Block 1\vspace{0.01cm}
        \includegraphics[scale=0.65]{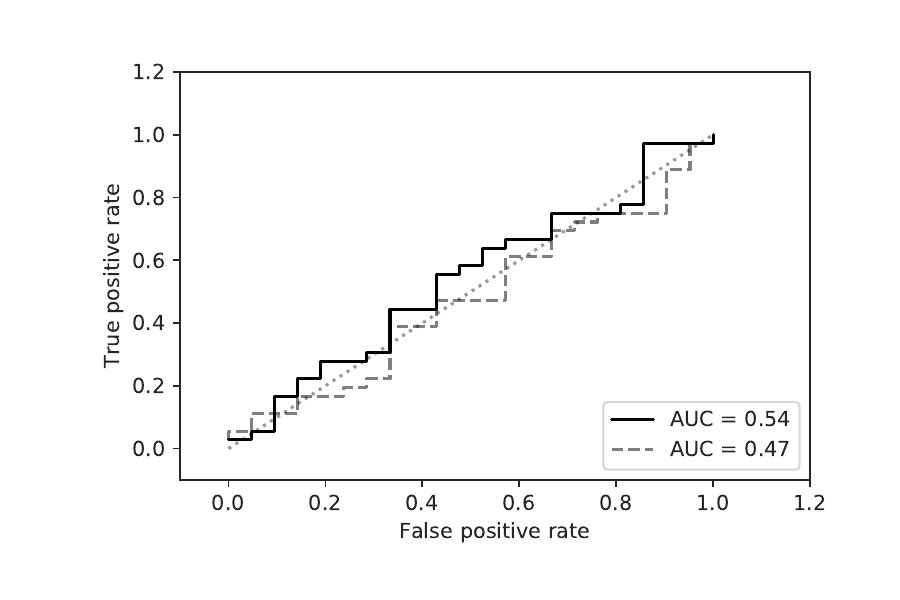}
\end{minipage} 
\hspace{0.72cm}
\begin{minipage}[b]{0.48\linewidth}
        \centering
        \qquad\qquad\qquad\quad Block 3\vspace{0.01cm}
        \includegraphics[scale=0.65]{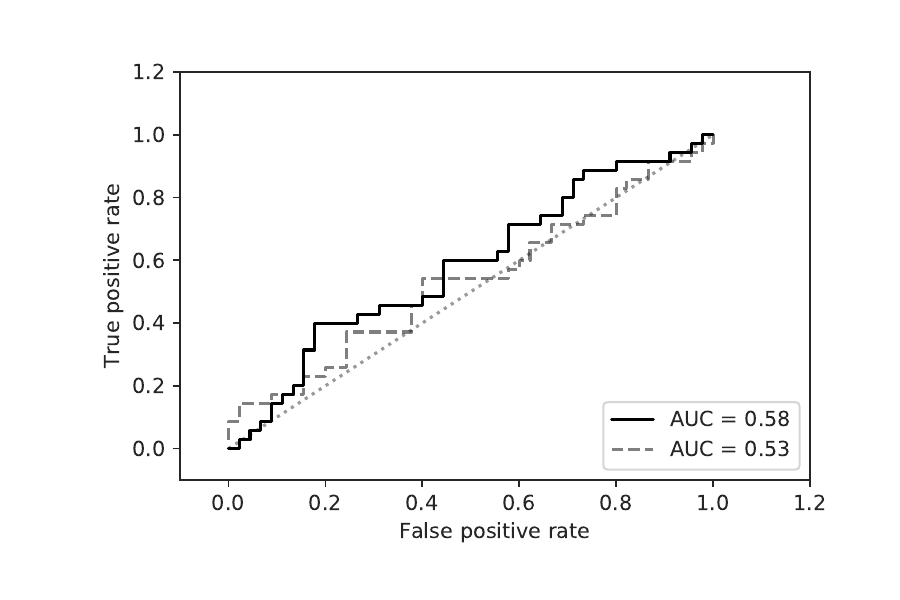}
\end{minipage} 
      \caption{ROC curves based on averaged probabilities for prediction blocks 1 and 3. \emph{Black} model including the latent player abilities, \emph{grey-dashed} baseline model, the \emph{dotted line} is the line $y=x$.} \label{rocfig}
\end{figure}

To form our predictions of whether over or under 2.5 goals are scored 
in a given game, we take each of our posterior draws (fitted using 
the previous block) and construct the $\theta_t$ of~\eqref{hier_pois} 
via \eqref{theta_h} and \eqref{theta_a} (baseline model), or 
\eqref{theta_h_ext} and \eqref{theta_a_ext} (including latent player 
abilities) for the games in the following block (our prediction 
block). Our prediction blocks are formed of games between teams we 
have already seen in the previous (fitting) blocks, hence prediction 
block 1 consists of 57~games (as we do not predict on promoted teams in 
this block), prediction blocks 2-4 are made up of 
80 games, with 60 games in prediction block~5. We use a 
predicted starting line-up from expert football analysts to determine $I_k^j$, 
the players who enter \eqref{fdeltah-pred} and \eqref{fdeltaa-pred}; 
these are human made starting line-ups, and do not come from a model, 
but they are usually quite accurate (86\% accuracy over the season) and 
vary little from the players who 
start a particular game. We then combine the $\theta_t$ for the home 
and away teams to give an overall scoring rate for each game, 
$\theta = \theta_h + \theta_a$, 
from which we calculate the probability of there being over 2.5 goals in 
the match. We average these probabilities across the posterior samples. 
ROC curves based on these averaged probabilities for prediction 
blocks 1 and 3 are presented in Figure~\ref{rocfig}. For clarity we also 
present the area under the curve (AUC) values for all prediction 
blocks in Table~\ref{tableroc}. The ROC curves for all prediction blocks 
are given in Appendix~\ref{roc_curves}.

\begin{table}
\caption{Area under the ROC curves based on averaged probabilities for each prediction block under both models.} \label{tableroc}
\centering
\begin{tabular}{l|ccccc}
\hline
\multicolumn{6}{c}{Area under the curve values}\\
\hline
& \multicolumn{5}{c}{Block}\\
Model & 1 & 2 & 3 & 4 & 5 \\
\hline
Baseline & 0.47 & 0.60 & 0.53 & 0.55 & 0.61 \\
Including latent player abilities & 0.54 & 0.65 & 0.58 & 0.68 & 0.62 \\
\hline
\end{tabular}
\end{table}

It is evident from both the figure and the table that including the 
latent player abilities in the model leads to a better predictive 
performance. We observe this increase across all blocks, although the 
difference between the models in block 5 is severely reduced compared 
to other blocks. The reasons for this reduction are twofold, the first 
being that given a near full season of data (2014/2015) on which we 
are predicting, the baseline model can reasonably accurately capture a 
team's attack and defence parameters better than it can towards the 
start of the season. Secondly the last block of a season tends to be 
more volatile as some teams try out younger players (who are not 
observed in the data previously), and others have increased motivation 
to score more goals to try and win games, for example, to avoid 
relegation. Whence, we observe similar behaviour under both models, as 
we observe less players in a starting line-up, moving the model 
including player abilities towards the baseline model. However, overall, 
we can conclude that the inclusion of the latent player abilities in 
the model results in a better predictive performance throughout the 
2014/2015 season.

    \subsubsection*{The over/under betting market} 

\begin{figure}
      \centering
      \includegraphics[scale=0.7]{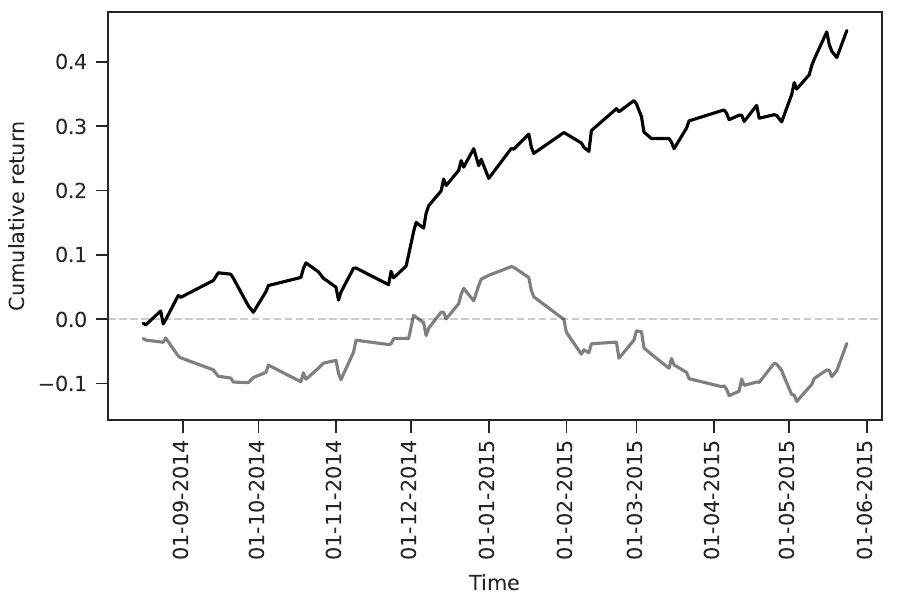}
      \caption{Cumulative return after betting on the over/under market for the 2014/2015 English Premier League season. \emph{Black} model including the latent player abilities, \emph{grey} baseline model.} \label{engoufig}
\end{figure}

As a final validation of the predictions made above, we consider both 
the baseline model and the model including the latent player abilities 
against the over/under betting market. We have odds data available to 
us (provided by Stratagem Technologies) for the 2014/2015 English 
Premier League season. Specifically, for 
each game we can bet on whether 2.5 goals will (over) or won't (under) 
be scored. A betting strategy, similar to one based on the Kelly 
Criterion, is used to determine which games are bet 
on. We are unfortunately unable to disclose the full details of the 
betting strategy used as it is connected to the business decisions 
of Stratagem Technologies. Recall that we do not make predictions for promoted teams 
in prediction block 1, and hence we do not bet on these games. 
The cumulative return over the season for both models is 
shown in Figure~\ref{engoufig}. As above, it is clearly evident that 
including the latent player abilities in the model leads to a better 
performance over the baseline model. It is expected that the baseline 
model will (in general) lead to a zero or slightly negative return, 
which is what we observe here, with the model fluctuating around zero. 
If we placed $\pounds 100$ on each bet, our return under the model 
including the latent player abilities would be $\pounds 4486.73$ whilst 
under the baseline model it would be $-\pounds 378.54$. Another indication 
that including the latent player abilities is worthwhile. From the above 
sections we can conclude that the inclusion of the latent player 
abilities in the model results in better predictions over the 
2014/2015 season.

\section{Discussion} \label{disc}

We have outlined a Bayesian model to establish player abilities. 
Our approach is computationally efficient and 
centres on variational inference methods. By adopting a Poisson model 
for occurrences of event types we are able to infer a player's ability 
for a multitude of event types, distinguishing between any two 
given players (even if they always play for the same team). 
These inferences are reasonably 
accurate and have close ties to reality, as seen in 
Section~\ref{ability}. Furthermore, our approach allows the 
visualisation of differences between players for a specific ability 
through the marginal posterior variational densities. We believe 
the rankings of players shown in this paper could lead to debate, and 
offer some evidence to the hotly debated questions of football.

We also extended the Bayesian hierarchical model of \cite{baio_2010} 
to include these latent player abilities. Through this model we 
captured a team's propensity to score goals, including a team's 
attacking ability, defensive ability and accounting for a home effect. 
We used output from this model to predict whether 2.5 goals would be 
scored in a game or not, observing an improvement in performance 
over the baseline model. A benefit of the prediction approach (and the 
block structure we implemented), is that it allowed us to see how our 
inferences about a player's ability evolved through time, explicitly 
highlighting what impact fringe players can have when they start getting 
regular playing time, for example, Kane in Section~\ref{prediction}.

\subsection*{Model assumptions and future work}

In constructing our model a number of assumptions have been made. Here we 
discuss the more prominent of these with the hope they will be improved upon 
in future work. Firstly, the choice of the event types (and their construction 
from other event types). Event types were created with the help of expert football 
analysts and it is possible that using different analysts would lead to 
different event types. However we believe, with the analysts suggestions, we 
have managed to capture the composition of a football game reasonably well. 
In terms of approach, we believe this to be a good starting point and a sensible 
route as all combinations of events would be too large a space to consider. The 
exact combination of these event types is an open question and one for further 
research. An independence is also assumed between event types when they are grouped. 
The groupings of the events come from expert football analysts, so we have made 
the assumption that they truly reflect the final ``grouped event,'' for example, 
AntiPass. The interactions between all these events is incredibly nuanced and 
unknown. To our knowledge no one has explored this problem of how all events 
on a football pitch interact. It is this aspect that makes football hard to fully 
predict, given its variability. If you could define all the interactions, you 
would be able to fully predict a match. The question of how to fully define all 
interactions is an interesting one, and one that would likely lead to years of 
interesting research. 

If a player is injured (or transfers to a different team during the year), they 
retain the same latent player abilities. We believe this to be a reasonable 
assumption but acknowledge some change in performance is likely if a player gets 
injured or moves team. We believe a change of performance would be quickly 
picked up by the model, as evidenced when we learn about new players, or players 
who start playing after being on the bench, see for example, Figure~\ref{goaltimefig}. 
An interesting extension may be to use a random walk on the variational 
parameters of a player when they change teams or return from injury.  

When making predictions we use a predicted starting line-up from expert football 
analysts. This obviously will not be available to all, and a possible future option 
to be explored is to use a predicted starting line-up based on an average 
line-up from a predetermined number of past games. Also, we have not removed 
penalties as we are interested in goals scored by a player (by any means), an 
extension to the model may be to model penalty goals separately from other goals. 
We felt this would be an over complication to the model in this scenario, given 
its relative complexity already. It is interesting to note that the top 3 players 
for Goal did not score one penalty between them, so it is possible that the 
penalty goals do not affect the result too greatly.

Finally, we have included home effects for all event types. In the future, when 
considering a larger number of event types we may find that some home effects 
for certain events are negligible, and as such may not be required within the 
model, leading to situations where the model can be made more parsimonious. 
Whilst it has not been explored within this paper, an interesting application 
going forward would be to consider adding a player from one team into another, 
creating the effect of transferring in a player. Team performance could then be 
predicted incorporating the added player's ability to see how a potential 
transfer target could affect results (or style of play). This of course could be 
done with multiple players, or entire fantasy squads, to help develop a better 
footballing strategy for the future.

In addition to addressing the above assumptions we plan three major ways of 
extending the current work. First, we 
intend to extend the variational approximation to allow for dependency 
among the latent abilities in the posterior.  Allowing for correlations 
in $q(\cdot)$ will let the model infer higher posterior variances resulting 
in a more robust ranking of players, and possibly improved predictive 
power for tasks such as providing probabilities on the number of goals 
in a future match.  From a modelling perspective, an extension is to 
let abilities change over time using a random walk across seasons and 
within seasons, which will be particularly useful when a substantial 
number of years of historical touch-by-touch data eventually becomes 
available. Finally, as the model gets applied to more competitions 
simultaneously it will be important to propose ways of scaling up 
the procedure. A topic worth investigating is how to best iteratively 
subsample the data for stochastic optimisation of the variational 
objective function. Of course, the approach proposed in this paper 
would also be applicable to other sports, and the authors believe such 
methods could easily be applied to basketball, hockey and rugby (with 
available data).

\section*{Acknowledgement}
The authors wish to thank Stratagem Technologies for providing the data 
and odds dataset, and especially to their expert football analysts for 
their help in defining event types, predicted starting line-ups and 
for interesting discussion and feedback throughout the project. 
Gavin A. Whitaker was funded, and Ricardo Silva partially funded, by 
KTP partnership KTP010441. Ricardo Silva and Ioannis Kosmidis have been 
supported by The Alan Turing Institute under the EPSRC grant EP/N510129/1.

\appendix
\section{Closed-form expression for the evidence lower bound} \label{app_elbo}

Recall that the log-likelihood to determine a player's ability 
for a specific event ($\Delta_i^e$) is given by 
\begin{equation}
\ell = \sum_{e\in E}\sum_{k=1}^K\sum_{j\in T_k}\sum_{i\in P_k^j} X_{i,k}^{e} \log{\left(\eta_{i,k}^{e}\tau_{i,k} \right)} - \eta_{i,k}^{e}\tau_{i,k} - \log{\left(X_{i,k}^{e}\,! \right)}, \label{app_llike}
\end{equation}
the evidence lower bound (ELBO) is 
\begin{align}
\elbo\left\{q\left(\Delta\right)\right\} & = \sum_{e\in E}\sum_{k=1}^K\sum_{j\in T_k}\sum_{i\in P_k^j} \expect_{q\left(\Delta_i^{e}\right)}\left[\log\left\{\pi\left(\Delta_i^{e}\right)\right\}\right] + \expect_{q\left(\Delta_i^{e}\right)}\left[\log\left\{\pi\left(x\vert\Delta_i^{e},\phi_i^{e},\psi\right)\right\}\right] \nonumber\\
&\qquad\qquad\qquad\qquad\quad- \expect_{q\left(\Delta_i^{e}\right)}\left[\log\left\{q\left(\Delta_i^{e}\vert\phi_i^{e}\right)\right\}\right],  \label{app_model_elbo}
\end{align}
and
\begin{equation}\label{app_qDelta_single}
q\left(\Delta_i^{e}\vert\phi_i^{e} \right)\sim N\left(\mu_{\Delta_i^{e}}, \sigma_{\Delta_i^{e}}^2 \right). 
\end{equation}
Here the ELBO is available in closed-form. 
Below we consider the terms in \eqref{app_model_elbo} on an 
individual basis to derive this closed-form. 

Let us begin by considering 
$\expect_{q\left(\Delta_i^{e}\right)}[\log\{q(\Delta_i^{e}\vert\phi_i^{e})\}]$. 
From \eqref{app_qDelta_single} we have 
\begin{equation*} 
\log\left\{q\left(\Delta_i^{e}\right)\right\} = -\frac{1}{2}\log\left(2\pi\sigma_{\Delta_i^{e}}^2 \right) - \frac{\left(\Delta_i^{e} - \mu_{\Delta_i^{e}} \right)^2}{2\sigma_{\Delta_i^{e}}^2}.
\end{equation*}
Taking expectations gives
\begin{align} 
\expect_{q\left(\Delta_i^{e}\right)}\left[\log\left\{q\left(\Delta_i^{e}\Big\vert\phi_i^{e}\right)\right\}\right] &= -\frac{1}{2}\log\left(2\pi\sigma_{\Delta_i^{e}}^2 \right)  \nonumber\\ 
& \qquad - \frac{1}{2\sigma_{\Delta_i^{e}}^2}\left[\expect\left\{\left(\Delta_i^{e}\right)^2\right\} - 2\expect\left(\Delta_i^{e}\right)\mu_{\Delta_i^{e}} + \left(\mu_{\Delta_i^{e}}\right)^2\right] \nonumber\\
&= -\frac{1}{2}\log\left(2\pi\sigma_{\Delta_i^{e}}^2 \right)  \nonumber\\ 
& \qquad - \frac{1}{2\sigma_{\Delta_i^{e}}^2}\left\{\sigma_{\Delta_i^{e}}^2 + \mu_{\Delta_i^{e}}^2 -2\mu_{\Delta_i^{e}}^2 + \mu_{\Delta_i^{e}}^2 \right\} \nonumber\\
&= -\frac{1}{2}\log\left(2\pi\sigma_{\Delta_i^{e}}^2 \right) -\frac{1}{2}, \label{qexpect}
\end{align}
which is the negative entropy of the Gaussian distribution. 

Let us now turn our attention to evaluating 
$\expect_{q\left(\Delta_i^{e}\right)}[\log\{\pi(\Delta_i^{e})\}]$, where 
$\Delta_i^{e}$ follows a $N(m,s^2)$ prior. Therefore
\begin{equation*} 
\log\left\{\pi\left(\Delta_i^{e}\right)\right\} =  -\frac{1}{2}\log\left(2\pi s^2\right) - \frac{\left(\Delta_i^{e} - m\right)^2}{2s^2}. 
\end{equation*}
Hence
\begin{align} 
\expect_{q\left(\Delta_i^{e}\right)}\left[\log\left\{\pi\left(\Delta_i^{e}\right)\right\}\right] &= -\frac{1}{2}\log\left(2\pi s^2 \right)  \nonumber\\     
& \qquad - \frac{1}{2s^2}\left[\expect\left\{\left(\Delta_i^{e}\right)^2\right\} - 2\expect\left(\Delta_i^{e}\right)m + m^2\right] \nonumber\\
&= -\frac{1}{2}\log\left(2\pi s^2 \right) - \frac{\sigma_{\Delta_i^{e}}^2 + \mu_{\Delta_i^{e}}^2 - 2m\mu_{\Delta_i^{e}} + m^2}{2s^2}. \label{pexpect}
\end{align}

Finally, let us consider $\expect_{q\left(\Delta_i^{e}\right)}[\log\{\pi(x\vert\Delta_i^{e},\phi_i^{e},\psi)\}]$. From \eqref{app_llike} we have
\[
\expect_{q\left(\Delta_i^{e}\right)}\left[\log\left\{\pi\left(x\Big\vert\Delta_i^{e},\phi_i^{e},\psi\right)\right\}\right]= \sum_{k=1}^K\underbrace{\expect\left\{X_{i,k}^{e} \log{\left(\eta_{i,k}^{e}\tau_{i,k} \right)} \right\}}_{\star} - \underbrace{\expect\left(\eta_{i,k}^{e}\tau_{i,k} \right)}_{\dagger} - \log{\left(X_{i,k}^{e}\,! \right)}.
\]
Evaluating $\star$ first gives
\begin{align*}
\star &= \expect\left[X_{i,k}^{e}\left\{\Delta_i^{e} 
                          + \tau_{i,k}\left(\lambda_1^e\sum_{i'\in P_k^j}\Delta_{i'}^{e}  
                          - \lambda_2^e\sum_{i'\in P_k^{T_k\setminus j}}\Delta_{i'}^{E\setminus e} \right)
                          + \left(\delta_{T_k^H,j}\right)\gamma^{\,e} + \log\left(\tau_{i,k}\right)\right\}\right] \nonumber\\ 
&= X_{i,k}^{e}\left\{\mu_{\Delta_i^{e}} + \tau_{i,k}\left(\lambda_1^e\sum_{i'\in P_k^j}\mu_{\Delta_{i'}^{e}} - \lambda_2^e\sum_{i'\in P_k^{T_k\setminus j}}\mu_{\Delta_{i'}^{E\setminus e}}\right) + \left(\delta_{T_k^H,j}\right)\gamma^{\,e}  + \log\left(\tau_{i,k}\right) \right\}.
\intertext{Turning to $\dagger$, we have}
\dagger&= \expect\left[\exp\left\{\Delta_i^{e} 
                          + \tau_{i,k}\left(\lambda_1^e\sum_{i'\in P_k^j}\Delta_{i'}^{e}  
                          - \lambda_2^e\sum_{i'\in P_k^{T_k\setminus j}}\Delta_{i'}^{E\setminus e} \right) 
                          + \left(\delta_{T_k^H,j}\right)\gamma^{\,e} \right\}\tau_{i,k} \right] \nonumber\\
&= \expect\left[\exp\left\{\left(1 + \tau_{i,k}\lambda_1^e\right)\Delta_i^{e}\right\}\exp\left(\tau_{i,k}\lambda_1^e\sum_{i'\in P_k^j\setminus i}\Delta_{i'}^{e}\right)\exp\left(-\tau_{i,k} \lambda_2^e\sum_{i'\in P_k^{T_k\setminus j}}\Delta_{i'}^{E\setminus e}\right)\exp\left\{\left(\delta_{T_k^H,j}\right)\gamma^{\,e}\right\}\tau_{i,k}\right]\nonumber\\
&= \exp\left\{\left(1 + \tau_{i,k}\lambda_1^e\right)\mu_{\Delta_i^{e}} + \frac{\left(1 + \tau_{i,k}\lambda_1^e\right)^2\sigma_{\Delta_i^{e}}^2}{2} \right\}
\exp\left\{\tau_{i,k}\lambda_1^e\sum_{i'\in P_k^j\setminus i}\mu_{\Delta_{i'}^{e}} + \frac{\left(\tau_{i,k}\lambda_1^e\right)^2\displaystyle\sum_{i'\in P_k^j\setminus i}\sigma_{\Delta_{i'}^{e}}^2}{2}\right\} \nonumber\\
&\qquad\qquad\times \exp\left\{-\tau_{i,k}\lambda_2^e\sum_{i'\in P_k^{T_k\setminus j}}\mu_{\Delta_{i'}^{E\setminus e}} + \frac{\left(-\tau_{i,k}\lambda_2^e\right)^2\displaystyle\sum_{i'\in P_k^{T_k\setminus j}}\sigma_{\Delta_{i'}^{E\setminus e}}^2 }{2} \right\} \exp\left\{\left(\delta_{T_k^H,j}\right)\gamma^{\,e} \right\}\tau_{i,k}.
\end{align*}
Therefore 
\begin{align}
\expect_{q\left(\Delta_i^{e}\right)}\left[\log\left\{\pi\left(x\Big\vert\Delta_i^{e},\phi_i^{e},\psi\right)\right\}\right]&= \sum_{k=1}^K\left(X_{i,k}^{e}\left\{\mu_{\Delta_i^{e}} + \tau_{i,k}\left(\lambda_1^e\sum_{i'\in P_k^j}\mu_{\Delta_{i'}^{e}} - \lambda_2^e\sum_{i'\in P_k^{T_k\setminus j}}\mu_{\Delta_{i'}^{E\setminus e}}\right) \right.\right.\nonumber \\
&\qquad\qquad\qquad \left. + \left(\delta_{T_k^H,j}\right)\gamma^{\,e} + \log\left(\tau_{i,k}\right) \right\} \nonumber\\
&\qquad - \left[\exp\left\{\left(1 + \tau_{i,k}\lambda_1^e\right)\mu_{\Delta_i^{e}} + \frac{\left(1 + \tau_{i,k}\lambda_1^e\right)^2\sigma_{\Delta_i^{e}}^2}{2} \right\} \right.\nonumber\\
&\left.\qquad\qquad\times\exp\left\{\tau_{i,k}\lambda_1^e\sum_{i'\in P_k^j\setminus i}\mu_{\Delta_{i'}^{e}} + \frac{\left(\tau_{i,k}\lambda_1^e\right)^2\displaystyle\sum_{i'\in P_k^j\setminus i}\sigma_{\Delta_{i'}^{e}}^2}{2}\right\} \right.\nonumber\\
&\left.\qquad\qquad\times \exp\left\{-\tau_{i,k}\lambda_2^e\sum_{i'\in P_k^{T_k\setminus j}}\mu_{\Delta_{i'}^{E\setminus e}} + \frac{\left(-\tau_{i,k}\lambda_2^e\right)^2\displaystyle\sum_{i'\in P_k^{T_k\setminus j}}\sigma_{\Delta_{i'}^{E\setminus e}}^2 }{2} \right\} \right.\nonumber\\
&\left.\left.\qquad\qquad\qquad\times\exp\left\{\left(\delta_{T_k^H,j}\right)\gamma^{\,e} \right\}\tau_{i,k}\right] - \log{\left\{X_{i,k}^{e}\,! \right\}}\right). \label{pxexpect}
\end{align}

Thus, the closed form of the ELBO \eqref{app_model_elbo} is obtained through 
a combination of \eqref{qexpect}--\eqref{pxexpect} whilst summing over 
$i$, $j$, $k$ and $e$, or explicitly
\begin{equation}
\elbo\left\{q\left(\Delta\right)\right\} = \sum_{e\in E}\sum_{j\in T_k}\sum_{i\in P_k^j} \textrm{\eqref{pexpect}} + \textrm{\eqref{pxexpect}} - \textrm{\eqref{qexpect}}.
\end{equation}

\section{Top 10 results} \label{app_top10}

In this section we detail top 10 rankings for a number of event types 
not presented in the Applications section---namely Shots, ShotStop, 
ChainEvents and AntiPass---which are presented in Tables~\ref{tableshots}, 
\ref{tableshotstop}, \ref{tablechain} and \ref{tableantipass} 
respectively. All 4 event types are of our own creation, made up of 
many other event types
\begin{itemize}
\item \textbf{Shots:} Goal, MissedShots, SavedShot, ShotOnPost.
\item \textbf{ShotStop:} Challenge, Claim, Interception, KeeperPickup, 
Punch, Save, Smother, Tackle.
\item \textbf{AntiPass:} BallRecovery, BlockedPass, Claim, Clearance, 
CornerAwarded, CrossNotClaimed, Interception, KeeperPickup, 
OffsideProvoked, Punch, Smother, Tackle.
\end{itemize}
Goal features in Shots, as a successful shot on target leads to a goal 
unless it becomes a SavedShot. ChainEvents is created by counting the 
number of instances a player is involved in the last 5 successful 
events leading to an event type contained within Shots, that is, the 
number of times a player is involved in a chain leading to a good 
attacking chance. The last 5 events were chosen as the length of the 
chain after discussion with expert football analysts, who thought that 
any further events back from the chance would have had little impact 
in creating it.

The top 10 for Shots consists entirely of strikers, the person seen as 
the main scorer of goals in a team, and thus, the person likely to have 
the most shots. The ranking appears sensible with the players 
heightened in our ranking (Aguero, Kane, Jovetic and A. Carroll) 
playing less time over the season due to injury, or mainly featuring 
as a substitute. The model suggests they took a large number of shots 
with the limited time they played. Suarez has an ability greater than 
any other player by a reasonable amount, which is expected given 
he had nearly 70 more shots than anyone else over the season. Over the 
2013/2014 English Premier League season Suarez was regarded as the 
best player, winning many awards, it is therefore unsurprising that he 
features highly in many of the top 10 rankings. 

The ranking for ShotStop is made up completely from goalkeepers, a 
natural conclusion given the event type. Lewis tops the ranking, 
although he only played 1 game and has a much larger standard deviation 
than anyone else. The goalkeepers for Fulham (Stockdale and 
Stekelenburg) played roughly half the season each, both stopping shots 
well (and at a similar level) when playing, for this reason they 
feature higher in our rankings than the observed order (determined by 
the total shots stopped over the season) suggests.

The top 10 for ChainEvents is similar in ways to that of GoalStop 
(presented in the main paper). It features a number of players with less 
playing time and therefore larger standard deviations. Whilst most of 
these players play a reasonable amount of time, from which to draw 
conclusions about their ability, the obvious outlier is Teixeira who 
played only 14 minutes (and has a very large standard deviation, 
comparatively). Again Suarez features highly in the rankings. The top 
10 contains the \emph{creative} players for each team, with that 
player for the top teams all featuring, Silva - Manchester City, 
Coutinho - Liverpool, Hazard - Chelsea. When we showed this ranking to 
expert football analysts there was a consensus that the ordering made 
sense (with the obvious exception of Teixeira).  

The ranking for AntiPass consists of both defenders and defensive 
midfielders, both types of player whose job it is to disrupt play. Alcaraz 
and Kallstrom have comparatively larger standard deviations, but the 
remainder of the ranking appears sensible. The difference between 
successive rankings is small, and there is some suggestion that it is 
easier to distinguish between player attacking ability than player 
defensive ability (GoalStop, ShotStop, AntiPass). This would agree with 
some in the football community who view attacking as an individual 
ability, whereas defending is more of a team ability. Overall, all 4 
of the rankings presented in this appendix appear largely 
sensible and agree with expert football analysts views.

\begin{table}
\caption{Top 10 shooters in the 2013/2014 English Premier League based on the 2.5\% quantile of the marginal posterior variational density for each player, $q(\Delta_i^{\textrm{Shots}})$.} \label{tableshots}
\centering
\footnotesize
\begin{tabular}{cllccccccc}
\hline
\multicolumn{10}{c}{Shots - top 10}\\
\hline
\multirow{2}{*}{Rank} & \multirow{2}{*}{Player} &  \multirow{2}{*}{Team} & 2.5\% & \multirow{2}{*}{Mean} & Standard & \multirow{2}{*}{Observed} & Observed & Rank & Time\\
 &  &  & quantile & & deviation &  & rank & difference & played \\
\hline
1 & Suarez 	& Liverpool &	1.426  	&	1.571  	&	0.074 	& 	181 &  1	& 0 & 3185\\ 	
2 & Aguero 	& Manchester City	& 	1.269 	& 	1.481 	& 	0.108 	& 	86 	 & 12 & +10 & 1616\\
3 & Dzeko   &	Manchester City &	1.220 	& 	1.414 	& 	0.099 	& 	103 	 &  5	& +2 & 2128\\
4 & Kane 	 & Tottenham Hotspur	& 	1.038 	& 	1.413 	& 	0.191 	& 	28 	 &  126	& +122 & 549 \\
5 & Bony 	& Swansea City	& 	1.036 	& 	1.225 	& 	0.096 	& 	108 	 & 3 	& -2 & 2644 \\
6 & Sturridge & Liverpool	& 	1.035 	& 	1.233 	& 	0.101 	& 	99 	 &  9	& +3 & 2414\\
7 & Jovetic 	& Manchester City	& 	1.027 	& 	1.441 	& 	0.211 	& 	23 	 & 150 	& +143 & 440\\
8 & Remy 	& Newcastle United	& 	0.998 	& 	1.206 	& 	0.106 	& 	90 	 &  11	& +3 & 2274 \\
9 & A. Carroll 	& West Ham United 	&	0.982 	& 	1.258 	& 	0.141 	& 	51 	 & 53	& +44 & 1200\\
10 & Jelavic & Everton/Hull	& 	 	0.978 	& 	1.210 	&  	0.118 	&  	72 & 18	& +8 & 1804\\
\hline
\end{tabular}
\end{table}

\begin{table}
\caption{Top 10 shot-stoppers in the 2013/2014 English Premier League based on the 2.5\% quantile of the marginal posterior variational density for each player, $q(\Delta_i^{\textrm{ShotStop}})$.} \label{tableshotstop}
\centering
\footnotesize
\begin{tabular}{cllccccccc}
\hline
\multicolumn{10}{c}{ShotStop - top 10}\\
\hline
\multirow{2}{*}{Rank} & \multirow{2}{*}{Player} &  \multirow{2}{*}{Team} & 2.5\% & \multirow{2}{*}{Mean} & Standard & \multirow{2}{*}{Observed} & Observed & Rank & Time\\
 &  &  & quantile & & deviation &  & rank & difference & played \\
\hline
1 & Lewis & Cardiff City	&	2.413 & 	2.823 & 	0.209 & 	23 & 369	& +368 & 98\\
2 & Mannone & Sunderland &	2.394 & 	2.485 & 	0.046 & 	453 & 8	& +6 & 2767 \\
3 & Ruddy & Norwich City &	2.312 & 	2.394 & 	0.042 & 	555 & 1	& -2 & 3679\\	
4 & Guzan & Aston Villa &	2.258 & 	2.343 & 	0.043 & 	512 & 2	& -2 & 3684\\
5 & Stockdale & Fulham	& 	2.225  &	2.343  &	0.060  &	267 & 19	& +14 & 1866\\ 
6 & Marshall 	& Cardiff City &	2.223  &	2.310  &	0.044 & 	497 & 3	& -3 & 3594\\ 	
7 & Stekelenburg & Fulham &	2.218  &	2.340 & 	0.062 & 	252 & 24	& +17 & 1790\\ 	
8 & Howard  & Everton &	2.218 & 	2.306 & 	0.045 & 	483 & 5	& -3 & 3575\\	
9 & Szczesny & Arsenal & 	2.199  &	2.286  &	0.045 & 	484 & 4	& -5 & 3594\\ 	
10 & Adrian & West Ham United	&	2.187 & 	2.306 & 	0.061 & 	262 & 20	& +10 & 1943\\	
\hline
\end{tabular}
\end{table}

\begin{table}
\caption{Top 10 players involved in the last 5 interactions leading to a chance in the 2013/2014 English Premier League based on the 2.5\% quantile of the marginal posterior variational density for each player, $q(\Delta_i^{\textrm{ChainEvents}})$.} \label{tablechain}
\centering
\footnotesize
\begin{tabular}{cllccccccc}
\hline
\multicolumn{10}{c}{ChainEvents - top 10}\\
\hline
\multirow{2}{*}{Rank} & \multirow{2}{*}{Player} &  \multirow{2}{*}{Team} & 2.5\% & \multirow{2}{*}{Mean} & Standard & \multirow{2}{*}{Observed} & Observed & Rank & Time\\
 &  &  & quantile & & deviation &  & rank & difference & played \\
\hline
1 & Teixeira & Liverpool &	2.419 &	3.291 &	0.444 &	5 & 474 & +473 &	14\\
2 & Suarez & Liverpool &	2.408 &	2.486 &	0.040 &	546  & 1 & -1 &	3185\\
3 & Eikrem & Cardiff City	& 	2.349 &	2.622 &	0.139 &	43 & 323 & +320 & 	229\\
4 & Jovetic & Manchester City &	2.303 &	2.525 &	0.113 &	72  & 250 & +246 &	440\\
5 & Silva 	& Manchester City &	2.299 &	2.394 &	0.049 &	372  & 5 & 0 &	2308\\
6 & Coutinho 	& Liverpool &	2.241 &	2.338 &	0.049 &	362  & 6 & 0 &	2473\\
7 & Taarabt 	& Fulham &	2.230 &	2.426 &	0.100 &	89  & 214 & +207 &	639\\
8 & Ramirez & Southampton &	2.218 &	2.415 &	0.101 &	92  & 208 & +200 &	601\\
9 & Aguero & Manchester City	& 	2.213 &	2.330 &	0.060 &	251  & 32 & +23 &	1616\\
10 & Hazard &	Chelsea & 	2.197 &	2.282 & 	0.043 & 	471  & 2 & -8 &	3100\\
\hline
\end{tabular}
\end{table}

\begin{table}
\caption{Top 10 anti-passers in the 2013/2014 English Premier League based on the 2.5\% quantile of the marginal posterior variational density for each player, $q(\Delta_i^{\textrm{AntiPass}})$.} \label{tableantipass}
\centering
\footnotesize
\begin{tabular}{cllccccccc}
\hline
\multicolumn{10}{c}{AntiPass - top 10}\\
\hline
\multirow{2}{*}{Rank} & \multirow{2}{*}{Player} &  \multirow{2}{*}{Team} & 2.5\% & \multirow{2}{*}{Mean} & Standard & \multirow{2}{*}{Observed} & Observed & Rank & Time\\
 &  &  & quantile & & deviation &  & rank & difference & played \\
\hline
1 & Alcaraz 	& Everton & 	2.864  & 	3.031  & 	0.085  & 	135  & 292 & +291 &	 	532\\
2 & Vidic 	 & Manchester United  & 	2.855  & 	2.940  & 	0.043  & 	520  & 35 & +33 &	 	2256\\
3 & Skrtel 	 &  Liverpool & 	2.814  & 	2.885  & 	0.036  & 	759  & 2 & -1 &	 	3468\\
4 & Kallstrom 	&  Arsenal  & 	2.793  & 	3.107  & 	0.160  & 	39   & 419 & +415 &	  	144\\
5 & Lovren 	   & Southampton  & 	2.789  & 	2.865  & 	0.039  & 	640  & 10 & +5 &	 	2993\\
6 & Koscielny &	Arsenal &	  2.783  & 	2.860  & 	0.039  & 	632  & 11 & +5 &	 	2980\\
7 & Mulumbu 	& West Bromwich Albion & 	2.778  & 	2.852  & 	0.038  & 	700  & 5 & -2 &	 	3319\\
8 & Azpilicueta & Chelsea	&	  2.769  & 	2.853  & 	0.043  & 	528  & 34 & +26 &	 	2522\\
9 & Jedinak 	  &  Crystal Palace  & 	2.763  & 	2.833  & 	0.036  & 	771  & 1 & -8 &	 	3651\\
10 & Fonte 	   &  Southampton & 	2.762  & 	2.835  & 	0.037  & 	709  & 4 & -6 &	 	3430\\	
\hline
\end{tabular}
\end{table}

\section{ROC curves based on averaged probabilities} \label{roc_curves}

\begin{figure}
\hspace{-1.2cm}
\begin{minipage}[b]{0.48\linewidth}
        \centering
        \qquad\qquad\qquad\quad Block 1\vspace{0.01cm}
        \includegraphics[scale=0.65]{both-roc-window1.pdf}
\end{minipage} 
\hspace{0.72cm}
\begin{minipage}[b]{0.48\linewidth}
        \centering
        \qquad\qquad\qquad\quad Block 2\vspace{0.01cm}
        \includegraphics[scale=0.65]{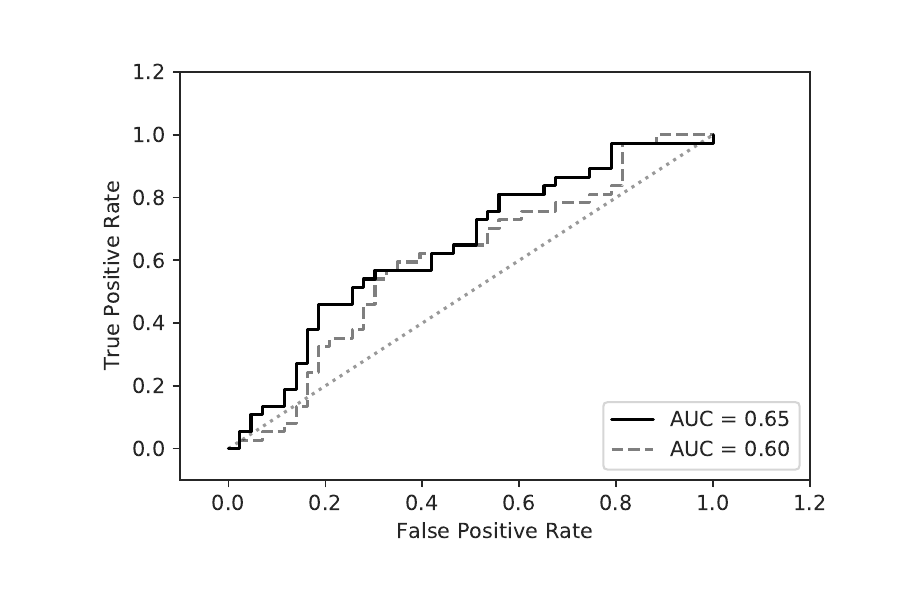}
\end{minipage} \vspace{0.2cm}\\
\hspace*{-1.2cm}
\begin{minipage}[b]{0.48\linewidth}
        \centering
        \qquad\qquad\qquad\quad Block 3\vspace{0.01cm}
        \includegraphics[scale=0.65]{both-roc-window3.pdf}
\end{minipage} 
\hspace{0.72cm}
\begin{minipage}[b]{0.48\linewidth}
        \centering
        \qquad\qquad\qquad\quad Block 4\vspace{0.01cm}
        \includegraphics[scale=0.65]{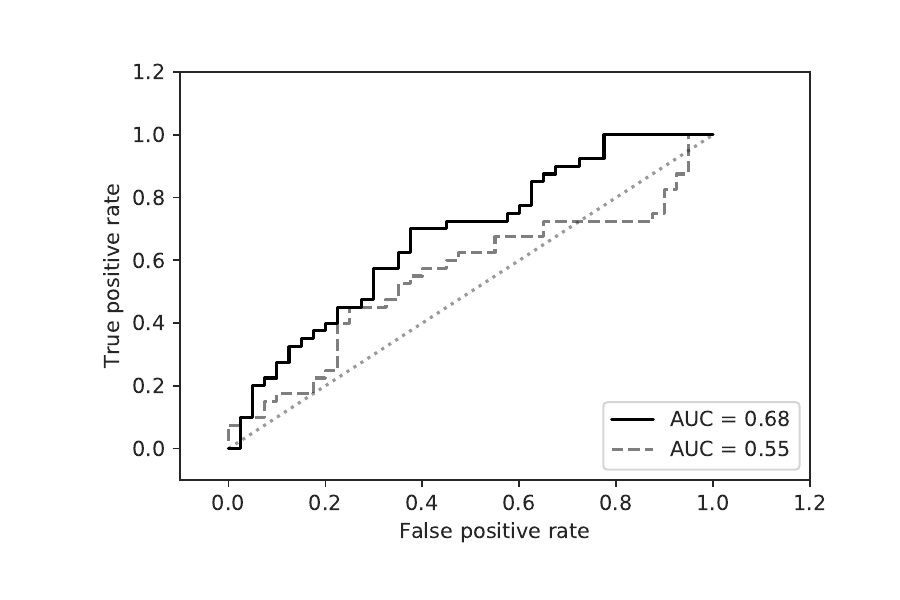}
\end{minipage} \vspace{0.2cm}\\
\hspace*{-1.2cm}
\begin{minipage}[b]{0.48\linewidth}
        \centering
        \qquad\qquad\qquad\quad Block 5\vspace{0.01cm}
        \includegraphics[scale=0.65]{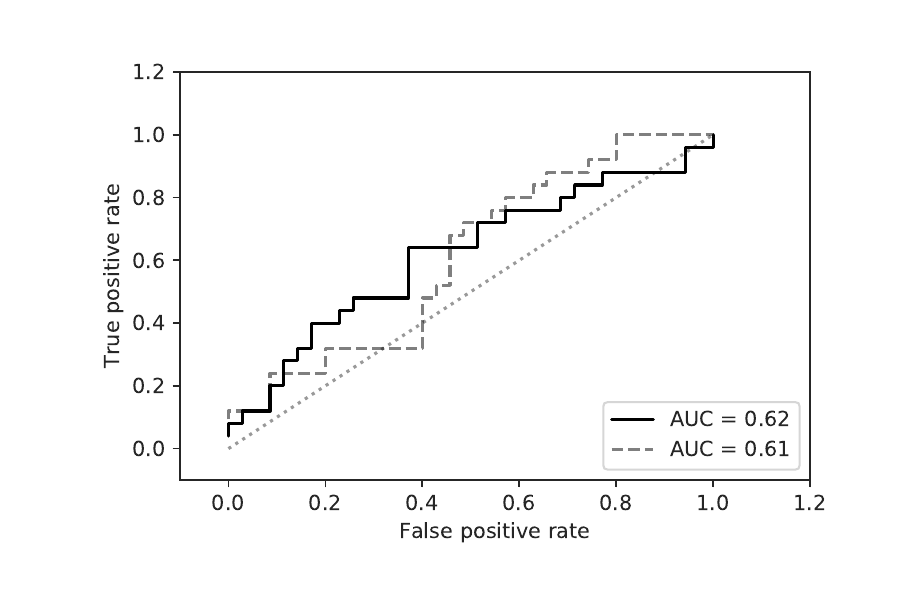}
\end{minipage}
      \caption{ROC curves based on averaged probabilities for each prediction block (see Section 4.2). \emph{Black}~model including the latent player abilities, \emph{grey-dashed} baseline model, the \emph{dotted line} is the line $y=x$.} \label{app_rocfig}
\end{figure}

ROC curves based on the averaged probabilities discussed in Section 4.2 
for each prediction block are presented in Figure~\ref{app_rocfig}.

\newpage
\bibliographystyle{apalike}
\bibliography{references}

\begin{thebibliography}{}

\bibitem[{AGR Analytics}, 2016]{AGR_2016}
{AGR Analytics} (2016).
\newblock Explaining and examining per 90.
\newblock
  \\\texttt{http://alexrathke.net/2016/07/explaining-and-examining-per-90}.

\bibitem[Aitchison and Ho, 1989]{aitchison_1989}
Aitchison, J. and Ho, C.~H. (1989).
\newblock {The multivariate Poisson-Log Normal distribution}.
\newblock {\em Biometrika}, 76(4):643--653.

\bibitem[Anderson and Sally, 2013]{anderson_2013}
Anderson, C. and Sally, D. (2013).
\newblock {\em The Numbers Game: Why Everything You Know About Football is
  Wrong}.
\newblock Penguin Books Limited.

\bibitem[Baio and Blangiardo, 2010]{baio_2010}
Baio, G. and Blangiardo, M. (2010).
\newblock Bayesian hierarchical model for the prediction of football results.
\newblock {\em Journal of Applied Statistics}, 37(2):253--264.

\bibitem[{BBC Business}, 2016]{bbc_2015}
{BBC Business} (2016).
\newblock Premier {L}eague in record \pounds 5.14bn {TV} rights deal.
\newblock \\\texttt{http://www.bbc.co.uk/news/business-31379128}.

\bibitem[Betfair, 2017]{betfair_2017}
Betfair (2017).
\newblock Over under 2.5 goals betting advice on {B}etfair.
\newblock
  \\\texttt{https://betting.betfair.com/over-under-25-goals-betting-advice-on-betfair.html}.

\bibitem[bet{HQ}, 2017]{bethq_2017}
bet{HQ} (2017).
\newblock Over/under goals betting.
\newblock
  \\\texttt{https://www.bethq.com/how-to-bet/articles/overunder-goals-betting}.

\bibitem[Bialkowski et~al., 2014]{bialkowski_2014}
Bialkowski, A., Lucey, P., Carr, P., Yue, Y., Sridharan, S., and Matthews, I.
  (2014).
\newblock Identifying team style in soccer using formations learned from
  spatiotemporal tracking data.
\newblock In {\em Data Mining Workshop ({ICDMW}), 2014 {IEEE} International
  Conference on}, pages 9--14. IEEE.

\bibitem[Bishop, 2006]{bishop_2006}
Bishop, C.~M. (2006).
\newblock {\em Pattern Recognition and Machine Learning}.
\newblock Information Science and Statistics. Springer.

\bibitem[Blei and Jordan, 2006]{blei_2006}
Blei, D.~M. and Jordan, M.~I. (2006).
\newblock Variational inference for {D}irichlet process mixtures.
\newblock {\em Bayesian Analysis}, 1(1):121--143.

\bibitem[Blei et~al., 2017]{blei_2017}
Blei, D.~M., Kucukelbir, A., and McAuliffe, J.~D. (2017).
\newblock Variational inference: A review for statisticians.
\newblock {\em Journal of the American Statistical Association},
  112(518):859--877.

\bibitem[Bojinov and Bornn, 2016]{bojinov_2016}
Bojinov, I. and Bornn, L. (2016).
\newblock The pressing game: Optimal defensive disruption in soccer.
\newblock In {\em {MIT} Sloan Sports Analytics Conference}. MIT SSAC.

\bibitem[Boshnakov et~al., 2017]{boshnakov_2017}
Boshnakov, G., Kharrat, T., and McHale, I.~G. (2017).
\newblock A bivariate {W}eibull count model for forecasting association
  football scores.
\newblock {\em International Journal of Forecasting}, 33(2):458--466.

\bibitem[Carbonetto and Stephens, 2012]{carbonetto_2012}
Carbonetto, P. and Stephens, M. (2012).
\newblock Scalable variational inference for {B}ayesian variable selection in
  regression, and its accuracy in genetic association studies.
\newblock {\em Bayesian Analysis}, 7(1):73--108.

\bibitem[Cave and Miller, 2016]{cave_2016}
Cave, A. and Miller, A. (2016).
\newblock Why football's {TV} deal is a game changer.
\newblock
  {\\\texttt{http://www.telegraph.co.uk/investing/business-of-sport/premier-league-investors/}}.

\bibitem[Chib and Winkelmann, 2001]{chib_2001}
Chib, S. and Winkelmann, R. (2001).
\newblock Markov chain {M}onte {C}arlo analysis of correlated count data.
\newblock {\em Journal of Business \& Economic Statistics}, 19(4):428--435.

\bibitem[Curley and Roeder, 2016]{curley_2016}
Curley, J.~P. and Roeder, O. (2016).
\newblock English soccer’s mysterious worldwide popularity.
\newblock
  {\texttt{https://contexts.org/articles/english-soccers-mysterious-worldwide-popularity/}}.

\bibitem[Deloitte, 2016]{deloitte_2016}
Deloitte (2016).
\newblock Deloitte's annual review of football finance.
\newblock
  \\{\scriptsize\texttt{https://www.deloitte.com/uk/en/pages/sports-business-group/articles/annual-review-of-football-finance.html}}.

\bibitem[Dixon and Coles, 1997]{dixon_1997}
Dixon, M.~J. and Coles, S.~G. (1997).
\newblock Modelling association football scores and inefficiencies in the
  football betting market.
\newblock {\em Journal of the Royal Statistical Society: Series C (Applied
  Statistics)}, 46(2):265--280.

\bibitem[Dixon and Robinson, 1998]{dixon_1998}
Dixon, M.~J. and Robinson, M. (1998).
\newblock A birth process model for association football matches.
\newblock {\em Journal of the Royal Statistical Society: Series D (The
  Statistician)}, 47(3):523--538.

\bibitem[Du et~al., 2009]{du_2009}
Du, L., Ren, L., Carin, L., and Dunson, D.~B. (2009).
\newblock A {B}ayesian model for simultaneous image clustering, annotation and
  object segmentation.
\newblock In {\em Advances in Neural Information Processing Systems}, pages
  486--494.

\bibitem[Franks et~al., 2015]{franks_2015}
Franks, A., Miller, A., Bornn, L., and Goldsberry, K. (2015).
\newblock Characterizing the spatial structure of defensive skill in
  professional basketball.
\newblock {\em The Annals of Applied Statistics}, 9(1):94--121.

\bibitem[Giordano et~al., 2018]{giordano_2018}
Giordano, R., Broderick, T., and Jordan, M.~I. (2018).
\newblock Covariances, robustness and variational bayes.
\newblock {\em The Journal of Machine Learning Research}, 19(1):1981--2029.

\bibitem[Groll et~al., 2018]{groll_2018}
Groll, A., Kneib, T., Mayr, A., and Schauberger, G. (2018).
\newblock On the dependency of soccer scores--a sparse bivariate {P}oisson
  model for the {UEFA E}uropean football championship 2016.
\newblock {\em Journal of Quantitative Analysis in Sports}, 14(2):65--79.

\bibitem[Herbrich et~al., 2007]{herbrich_2007}
Herbrich, R., Minka, T., and Graepel, T. (2007).
\newblock Trueskill\texttrademark : A {B}ayesian skill rating system.
\newblock In Sch\"{o}lkopf, B., Platt, J.~C., and Hoffman, T., editors, {\em
  Advances in Neural Information Processing Systems 19}, pages 569--576. MIT
  Press.

\bibitem[Jordan et~al., 1999]{jordan_1999}
Jordan, M.~I., Ghahramani, Z., Jaakkola, T.~S., and Saul, L.~K. (1999).
\newblock An introduction to variational methods for graphical models.
\newblock {\em Machine Learning}, 37(2):183--233.

\bibitem[Joseph et~al., 2006]{joseph_2006}
Joseph, A., Fenton, N.~E., and Neil, M. (2006).
\newblock Predicting football results using {B}ayesian nets and other machine
  learning techniques.
\newblock {\em Knowledge-Based Systems}, 19(7):544--553.

\bibitem[Karlis and Ntzoufras, 2000]{karlis_2000}
Karlis, D. and Ntzoufras, I. (2000).
\newblock On modelling soccer data.
\newblock {\em Student}, 3:229--245.

\bibitem[Karlis and Ntzoufras, 2003]{karlis_2003}
Karlis, D. and Ntzoufras, I. (2003).
\newblock Analysis of sports data by using bivariate {P}oisson models.
\newblock {\em Journal of the Royal Statistical Society: Series D (The
  Statistician)}, 52(3):381--393.

\bibitem[Karlis and Ntzoufras, 2009]{karlis_2009}
Karlis, D. and Ntzoufras, I. (2009).
\newblock Bayesian modelling of football outcomes: using the {S}kellam's
  distribution for the goal difference.
\newblock {\em IMA Journal of Management Mathematics}, 20(2):133--145.

\bibitem[Kitani et~al., 2011]{kitani_2011}
Kitani, K.~M., Okabe, T., Sato, Y., and Sugimoto, A. (2011).
\newblock Fast unsupervised ego-action learning for first-person sports videos.
\newblock In {\em Computer Vision and Pattern Recognition (CVPR), IEEE
  Conference on}, pages 3241--3248.

\bibitem[Kucukelbir et~al., 2017]{kucukelbir_2016}
Kucukelbir, A., Tran, D., Ranganath, R., Gelman, A., and Blei, D.~M. (2017).
\newblock Automatic differentiation variational inference.
\newblock {\em The Journal of Machine Learning Research}, 18(1):430--474.

\bibitem[Lee, 1997]{lee_1997}
Lee, A.~J. (1997).
\newblock Modeling scores in the {P}remier {L}eague: {I}s {M}anchester {U}nited
  really the best?
\newblock {\em Chance}, 10(1):15--19.

\bibitem[Lucey et~al., 2013]{lucey_2013}
Lucey, P., Oliver, D., Carr, P., Roth, J., and Matthews, I. (2013).
\newblock Assessing team strategy using spatiotemporal data.
\newblock In {\em Proceedings of the 19th ACM SIGKDD International Conference
  on Knowledge Discovery and Data Mining}, pages 1366--1374. ACM.

\bibitem[Maclaurin et~al., 2015]{maclaurin_2015}
Maclaurin, D., Duvenaud, D., and Adams, R.~P. (2015).
\newblock Autograd: Effortless gradients in numpy.
\newblock In {\em ICML 2015 AutoML Workshop}.

\bibitem[Maher, 1982]{maher_1982}
Maher, M.~J. (1982).
\newblock Modelling association football scores.
\newblock {\em Statistica Neerlandica}, 36(3):109--118.

\bibitem[McHale et~al., 2012]{mchale_2012}
McHale, I.~G., Scarf, P.~A., and Folker, D.~E. (2012).
\newblock On the development of a soccer player performance rating system for
  the {E}nglish {P}remier {L}eague.
\newblock {\em Interfaces}, 42(4):339--351.

\bibitem[McHale and Szczepa{\'n}ski, 2014]{mchale_2014}
McHale, I.~G. and Szczepa{\'n}ski, {\L}. (2014).
\newblock A mixed effects model for identifying goal scoring ability of
  footballers.
\newblock {\em Journal of the Royal Statistical Society: Series A (Statistics
  in Society)}, 177(2):397--417.

\bibitem[Minka et~al., 2018]{minka_2018}
Minka, T., Cleven, R., and Zaykov, Y. (2018).
\newblock {TrueSkill 2: An improved Bayesian skill rating system}.
\newblock Technical Report MSR-TR-2018-8, Microsoft.

\bibitem[Raj et~al., 2014]{raj_2014}
Raj, A., Stephens, M., and Pritchard, J.~K. (2014).
\newblock {fastSTRUCTURE: Variational inference of population structure in
  large SNP data sets}.
\newblock {\em Genetics}, 197(2):573--589.

\bibitem[Reep et~al., 1971]{reep_1971}
Reep, C., Pollard, R., and Benjamin, B. (1971).
\newblock Skill and chance in ball games.
\newblock {\em Journal of the Royal Statistical Society. Series A (General)},
  pages 623--629.

\bibitem[Ruiz and Perez-Cruz, 2015]{ruiz_2015}
Ruiz, F. J.~R. and Perez-Cruz, F. (2015).
\newblock A generative model for predicting outcomes in college basketball.
\newblock {\em Journal of Quantitative Analysis in Sports}, 11(1):39--52.

\bibitem[Rumsby, 2016]{rumsby_2016}
Rumsby, B. (2016).
\newblock Premier {L}eague clubs to share \pounds 8.3 billion {TV} windfall.
\newblock
  \\{\scriptsize\texttt{http://www.telegraph.co.uk/sport/football/12141415/Premier-League-clubs-to-share-8.3-billion-TV-windfall.html}}.

\bibitem[Saul and Jordan, 1996]{saul_1996}
Saul, L. and Jordan, M.~I. (1996).
\newblock Exploiting tractable substructures in intractable networks.
\newblock In {\em Advances in Neural Information Processing Systems 8}, pages
  486--492. MIT Press.

\bibitem[{SPORTINGINDEX}, 2017]{Sportingindex_2017}
{SPORTINGINDEX} (2017).
\newblock Most popular spread betting markets.
\newblock
  \\\texttt{https://www.sportingindex.com/\#/o/learn/training-centre/most-popular-markets}.

\bibitem[{Stan Development Team}, 2016]{stan_2016}
{Stan Development Team} (2016).
\newblock {PyStan: the Python interface to Stan, version 2.15.0.0}.

\bibitem[Sudderth and Jordan, 2009]{sudderth_2009}
Sudderth, E.~B. and Jordan, M.~I. (2009).
\newblock Shared segmentation of natural scenes using dependent {Pitman-Yor}
  processes.
\newblock In {\em Advances in Neural Information Processing Systems}, pages
  1585--1592.

\bibitem[Tunaru, 2002]{tunaru_2002}
Tunaru, R. (2002).
\newblock Hierarchical {B}ayesian models for multiple count data.
\newblock {\em Austrian Journal of statistics}, 31(3):221--229.

\bibitem[Wainwright and Jordan, 2008]{wainwright_2008}
Wainwright, M.~J. and Jordan, M.~I. (2008).
\newblock Graphical models, exponential families, and variational inference.
\newblock {\em Foundations and Trends in Machine Learning}, 1(1-2):1--305.

\bibitem[Whitaker et~al., 2018]{whitaker_2018}
Whitaker, G.~A., Silva, R., and Edwards, D. (2018).
\newblock Visualizing a team's goal chances in soccer from attacking events: A
  {B}ayesian inference approach.
\newblock {\em Big Data}, 6(4):271--290.

\bibitem[Yueh, 2014]{yueh_2014}
Yueh, L. (2014).
\newblock Exporting football. {W}hy does the world love the {E}nglish {P}remier
  {L}eague?
\newblock \texttt{http://www.bbc.co.uk/news/business-27369580}.

\end{thebibliography}

\end{document}